\documentclass[aps,prd,10pt,twocolumn,superscriptaddress,floatfix,nofootinbib,amsmath,amssymb,altaffilletter,preprintnumbers]{revtex4-1}
\usepackage[colorlinks=true,pdfstartview=FitV,breaklinks=true]{hyperref}

\usepackage[utf8]{inputenc}
\usepackage{graphicx}
\usepackage{times}
\usepackage{booktabs}
\usepackage{tabularx}
\usepackage{textcomp}
\usepackage{xspace}
\usepackage{amsmath}
\usepackage{upgreek}
\usepackage{amsfonts}
\usepackage{amssymb}
\usepackage{mdframed}
\usepackage[dvipsnames,table]{xcolor}
\newcolumntype{Y}{>{\centering\arraybackslash}X}
\newcommand{\ra}[1]{\renewcommand{\arraystretch}{#1}}

\hypersetup{urlcolor=BlueViolet,
	    citecolor=Plum,
	    linkcolor=PineGreen}
\usepackage[official]{eurosym}
\definecolor{lightgray}{rgb}{0.9,0.9,0.9}	    
\definecolor{green}{rgb}{0,0.5,0}
\definecolor{red}{rgb}{1,0,0}
\definecolor{blue}{rgb}{0,0,0.5}

\newcommand\codo[1]{{\sc #1}}
\newcommand{\sech}{\operatorname{sech}}
\newcommand{\n}{{\ensuremath{\emph{n}}}\xspace}

\begin{document}

\title{Simulations of axion-like particles in the post-inflationary scenario}

\author{Ciaran A. J. O'Hare}
\affiliation{School of Physics, The University of Sydney, and ARC Centre of Excellence for Dark Matter Particle Physics, NSW 2006, Australia}

\author{Giovanni Pierobon}\thanks{Corresponding author}\email{g.pierobon@unsw.edu.au}
\affiliation{School of Physics, The University of New South Wales, Sydney NSW 2052, Australia}

\author{Javier Redondo}
\affiliation{Departamento de Fisica Teorica, Universidad de Zaragoza, 50009 Zaragoza, Spain}
\affiliation{Max-Planck-Institut f\"ur Physik (Werner-Heisenberg-Institut), F\"ohringer Ring 6, 80805 M\"unchen, Germany}

\author{Yvonne Y. Y. Wong}
\affiliation{School of Physics, The University of New South Wales, Sydney NSW 2052, Australia}

\preprint{CPPC-2021-09}

\date{\today}
\smallskip
\begin{abstract}
Axions and axion-like particles (ALPs) are some of the most popular candidates for dark matter, with several viable production scenarios that make different predictions. In the scenario in which the axion is born after inflation, its field develops significant inhomogeneity and evolves in a highly nonlinear fashion. Understanding the eventual abundance and distribution of axionic dark matter in this scenario therefore requires dedicated numerical simulations. So far the community has focused its efforts on simulations of the QCD axion, a model that predicts a specific temperature dependence for the axion mass. Here, we go beyond the QCD axion, and perform a suite of simulations on lattice sizes of $3072^3$, over a range of possible temperature dependencies labelled by a power-law index $\n \in [0,6]$.  We study the complex dynamics of the axion field, including the scaling of cosmic strings and domain walls, the spectrum of non-relativistic axions, the lifetime and internal structure of axitons, and the seeds of miniclusters. In particular, we quantify how much the string-wall network contributes to the dark matter abundance as a function of how quickly the axion mass grows. We find that a temperature-independent model produces 25\% more dark matter than the standard misalignment calculation. In contrast to this generic ALP, QCD axion models are almost six times less efficient at producing dark matter. Given the flourishing experimental campaign to search for ALPs, these results have potentially wide implications for direct and indirect searches.
\end{abstract}

\maketitle

\section{Introduction}
Although originally proposed to solve the strong-$CP$ problem of quantum chromodynamics (QCD)~\cite{Peccei:1977hh,Kim:2008hd,DiLuzio:2020wdo}, the \emph{axion} is now currently one of the most compelling explanations for why the Universe is filled with dark matter (DM)~\cite{Bertone:2016nfn,Preskill:1982cy,Abbott:1982af,Dine:1981rt,Dine:1982ah,Turner:1983he,Turner:1985si}. So popular is the axion as a candidate for DM that an entire class of related axion-like particles (ALPs) has also sprung up through the years~\cite{Marsh:2015xka}.  These ALPs are {\it a priori} unconnected to QCD, but
may nonetheless provide a window on high-energy physics far beyond the reach of colliders~\cite{Masso:1995tw, Masso:2002ip, Ringwald:2012hr, Ringwald:2012cu, Arvanitaki:2009fg, Cicoli:2012sz, Jaeckel:2010ni}. For this reason, and their potential role as DM, ALPs are now a subject of growing theoretical~\cite{Choi:2020rgn} and experimental~\cite{Irastorza:2018dyq} interest.

The axion or ALP field is produced at an energy scale $f_a$ when a symmetry---of which the axion or the ALP is the associated pseudo-Goldstone boson---is broken. This scale is a free parameter, but we can generally classify the symmetry breaking as having occurred either before or after inflation---a distinction that has important consequences for the subsequent population of axions.%
\footnote{Unless specifically designated as a QCD axion, we shall use the terms axions and ALPs interchangeably.}
In the so-called \emph{pre-inflationary} scenario, inflation blows up a tiny region of the initially randomised axion field, and for a sensible range of initial field values, the  observed DM abundance can be straightforwardly matched to theoretical predictions to identify a range of responsible axion masses. 

Conversely, in the $N_{\rm DW}=1$ \emph{post-inflationary} scenario, once the axion has been produced, the Universe is instead filled with an ensemble of random angular field values drawn from $[-\pi, \pi]$ that should in principle approach a predictable average. However, these random angles also make the axion field highly inhomogeneous and lead to the emergence of topological defects~\cite{Kibble:1976sj,Kibble:1982dd,Vilenkin:1982ks}. As the Universe expands, the defect network relaxes and eventually collapses, but not before leaving a substantial imprint on the axion population~\cite{Vilenkin:1986ku,Davis:1986xc,Harari:1987ht,Davis:1985pt,Battye:1993jv}.  Consequently, the distribution of axions in the post-inflationary scenario is rather complex, and its investigation requires numerical simulations. Only very recently have simulations been designed with the sophistication needed to study the cosmological axion~\cite{Hiramatsu:2012gg,Kawasaki:2014sqa,Fleury:2015aca,Klaer:2017ond,Vaquero:2018tib,Buschmann:2019icd,Gorghetto:2018myk,Gorghetto:2020qws,Buschmann:2021sdq,Hindmarsh:2019csc,Hindmarsh:2021vih}, or more specifically, to estimate the DM axion mass from the observed abundance with some certainty (see, e.g., Ref.~\cite{Hoof:2021jft} for a recent discussion).

But the need for cosmological axion simulations goes beyond predicting the DM abundance.
Simulations are also crucial for the investigation of the phase-space distribution of the axions produced. Long after topological defects have decayed away and the axions have free-streamed until they are non-relativistic, regions of high density will eventually collapse under gravity to form structures (see, e.g., Ref.~\cite{Niemeyer:2019aqm} for a recent review). The overdensities that seed these structures are remnants of dense field configurations known as \emph{axitons}---also known as oscillons or pseudo-breathers---which appear briefly while the axion mass is still growing to its present-day value~\cite{Vaquero:2018tib,Buschmann:2019icd}. By around matter-radiation equality these overdensities will have collapsed into small gravitationally bound clumps of axions called \emph{miniclusters}~\cite{Hogan:1988mp,Kolb:1993zz,Kolb:1993hw,Zurek:2006sy,Kolb:1994fi,Hardy:2016mns,Davidson:2016uok,Enander:2017ogx}, with masses around that of small asteroids.

Axion miniclusters are likely to survive to the present time (see, e.g., Refs.~\cite{Eggemeier:2019khm,Arvanitaki:2019rax,Ellis:2020gtq,Blinov:2019jqc,Xiao:2021nkb}), so it is certainly possible that the majority of the axions making up galactic halos could still be bound up inside of them. This degree of small-scale DM substructure is a unique prediction of the post-inflationary axionic DM scenario, and has the potential to both hinder direct experimental searches~\cite{Tinyakov:2015cgg,Dokuchaev,OHare:2017yze,Knirck:2018knd,Kavanagh:2020gcy}, as well as facilitate entirely new indirect ones~\cite{Kolb:1995bu,Tkachev:2014dpa,Pshirkov:2016bjr,Fairbairn:2017sil,Katz:2018zrn,Dai:2019lud,Croon:2020wpr,Edwards:2020afl}. Simulations are therefore crucial not only for predicting where to find the axion mass, but also to inform us what kinds of searches are even possible.

One of the key theoretical inputs to cosmological axion simulations is the temperature dependence of the axion mass, usually modelled as $m^2_a \propto T^{-\n}$. In QCD axion models, this mass originates from the explicit breaking of the Peccei-Quinn (PQ) symmetry due to its colour anomaly, and as the temperature drops, its dependence on $T$ can be predicted with reasonable confidence from theoretical models~\cite{Wantz:2009it} or lattice calculations~\cite{Borsanyi:2016ksw,GrillidiCortona:2015jxo}. However, all computational methods have inherent uncertainties, and there could exist alternative axion models that have altogether different temperature dependencies. It may therefore be preferable  to consider $\n$ as a restrained but ultimately free parameter. Moreover, by treating $\n$ as such, we can gain further physical insight into the role played by the axion mass growth rate in governing the complex evolution of the field. One interesting example is the era in which the field's energy density fluctuations are dominated by axitons---objects that are supported in an expanding background precisely by the growing axion mass. Since axitons lay down perturbations on the scales of the resulting miniclusters~\cite{Vaquero:2018tib}, we expect any $\n$-dependence to leave an imprint in the axion distribution surviving well beyond the final times of our simulations.

\begin{figure}[t]
    \centering
    \includegraphics[width=0.99\columnwidth]{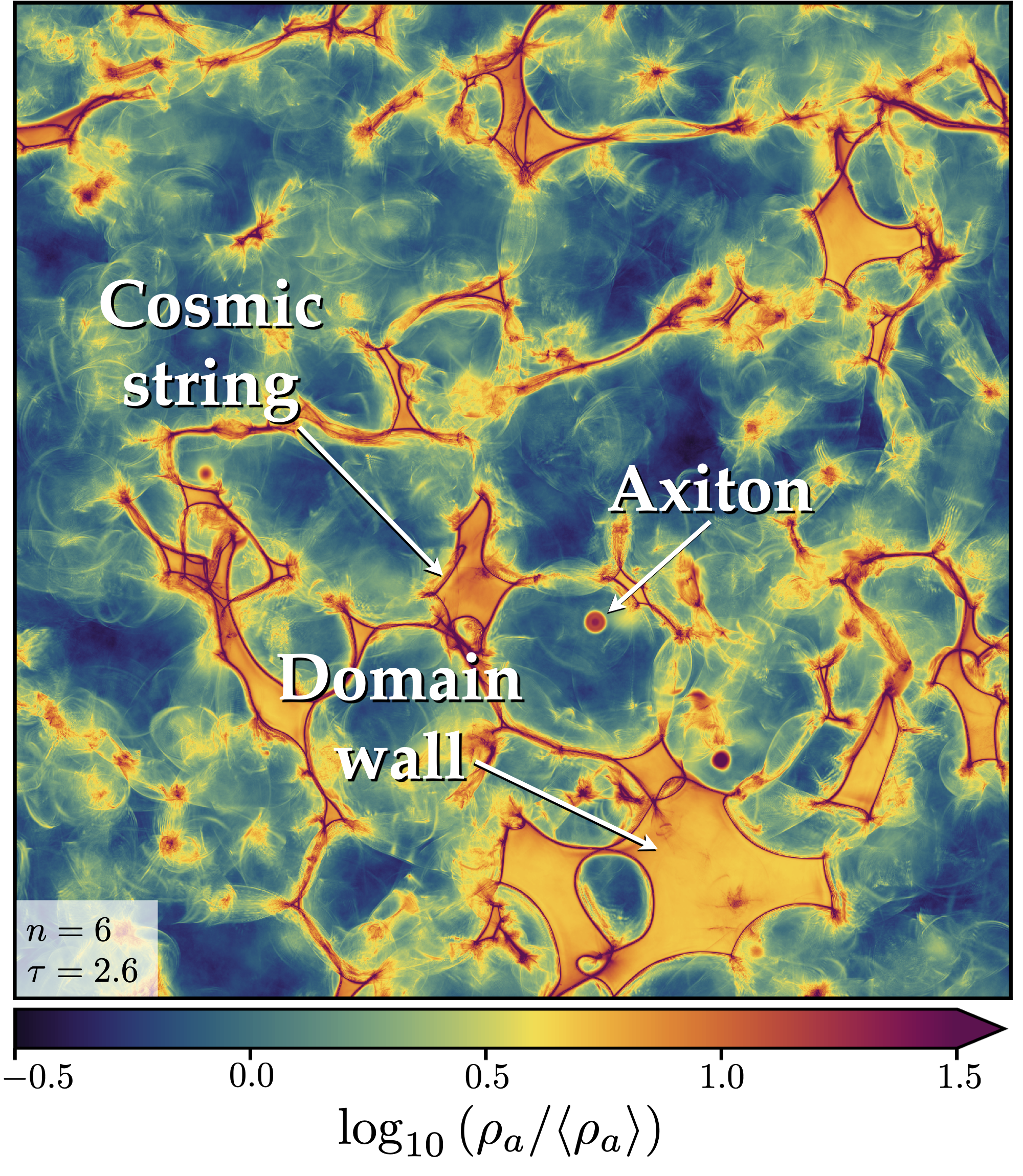}
    \caption{A 3D-to-2D projection of the axion energy density around the time the string-wall network is almost fully collapsed. The field is shown here at the dimensionless simulation time $\tau\equiv R/R_1 = 2.6$, where $R_1$ is the cosmological scale factor at the time the axion field begins to oscillate around the minimum of its potential. We highlight the presence of the three main types of object to be discussed: cosmic strings, domain walls, and spherical overdensities called axitons. This particular snapshot is from a simulation where we have chosen the axion mass to grow with temperature as $m^2_a\propto T^{-6}$, and at a time when the comoving size of the box contains $\sim 60$ comoving axion Compton wavelengths, defined as $\tau\lambda_{c}=2\pi/m_a$. Additional visualisations can be found in \href{https://www.youtube.com/channel/UCqosWwoOKrdmNKBbIwVUuGg/videos}{yt/jaxions}.
    }
    \label{fig:introfig}
\end{figure}

In this paper, we build upon and extend previous works by performing a suite of seventy simulations encompassing $\n \in [0,6]$, using the public code~\codo{jaxions}\footnote{\href{https://github.com/veintemillas/jaxions}{https://github.com/veintemillas/jaxions}.} developed originally in Ref.~\cite{Vaquero:2018tib}. These include the first large-scale lattice simulations of a generic axion-like particle model (i.e., the case $\n =0$). We use these simulations to systematically study the role played by the axion mass parameterisation in dictating the energy spectrum of axions, the population and properties of axitons and miniclusters, as well as the eventual abundance of DM. A snapshot of the projected energy density contained in the axion field for the case $\n=6$ can be seen in Fig.~\ref{fig:introfig}, where we highlight several of the important classes of object that much of our discussion centres around.

The paper is structured as follows.  We begin in Sec.~\ref{sec:simulation} with an outline of the simulation setup, including details on the model, units, and initial conditions. Section~\ref{sec:dynamics} gives a general description of the main stages of the field's evolution seen in our simulations. We then describe and compare over several sections our various simulation outputs for different values of~$\n$. We present first an analysis of the properties of axitons in Sec.~\ref{sec:axitons}, followed by the shape of the axion power spectrum in Sec.~\ref{sec:specshape} and the DM yield in Sec.~\ref{sec:darkmatter}, before ending with a discussion on the seeds of axion miniclusters in Sec.~\ref{sec:gravity}. Section~\ref{sec:conclusions} contains our conclusions.

\section{The simulation}\label{sec:simulation}

 The basic scheme of our simulations centres around a complex scalar field, discretised on a regular lattice and evolved through cosmic time according to its equation of motion. We adopt a generic potential that dictates the evolution of both the \emph{radial} part of the field---entailing the breaking of the PQ symmetry and the appearance of cosmic strings---as well as a term that describes the growth of the mass of the \emph{angular} component of the field, and the subsequent appearance of domain walls and axitons. 

\subsection{Model}
The action for the complex scalar field $\phi$, usually referred to as the PQ field in the context of axion physics, is
\begin{equation}
\label{eq:action}
    \mathcal{S}=\int d^4x \sqrt{-g}\bigg[\vert\partial \phi\vert^2-V(\phi)\bigg], \;\;\;\;\;\;\; \phi(x)=\vert\phi\vert e^{i\theta(x)}.
\end{equation}  
Here, $g \equiv {\rm det} \left[g_{\mu \nu} \right]$, where $g_{\mu\nu}$ is taken to be the flat Friedmann-Lema\^{\i}tre-Robertson-Walker (FLRW) metric defined via ${{\rm d}s^2 = g_{\nu \mu} {\rm d} x^\mu {\rm d}x^\nu = {\rm d} t^2 - R^2(t) {\rm d} x^i {\rm d}x_i}$ for the cosmic time $t$ and comoving spatial coordinates $x^i$, with scale factor~$R$.
 The axion field $a(x) \equiv \theta(x)f_a$ is identified with the phase of $\phi$ in units of $f_a$, while the massive radial mode $\vert\phi\vert$ is often called the saxion field. The potential $V(\phi)$ can be written as the sum of two terms,
\begin{equation}\label{eq:fullpot} 
V=V_s(\vert\phi\vert)+V_a(\theta)=\frac{\lambda_{\phi}}{8}(\vert\phi\vert^2-f_a^2)^2+\chi(T)(1-\cos\theta).
\end{equation}
The saxion potential $V_s(\vert\phi\vert)$ is responsible for the spontaneous symmetry breaking and gives the complex field its vacuum expectation value (vev) $\langle\phi\rangle=f_a$. The mass of the saxion field can be simply read off from the potential as $m_s^2=\lambda_{\phi}f^2_a$. 

The second term, $V_a(\theta)$, is the axion potential, whose temperature dependence is provided by a function called the topological susceptibility $\chi(T)$. In QCD axion models this potential originates from the breaking of the anomalous PQ symmetry via QCD instantons, and its growth in time is due to the temperature dependence of those instantons. We can therefore expect the axion potential to become important around the QCD scale $\Lambda_{\rm QCD}$. At high temperatures ($T > \Lambda_{\rm QCD}$) the functional dependence of $\chi(T)$ can be parameterised as
\begin{equation}
    \chi(T)=\chi_0\left(\frac{T}{T^{\rm QCD}_{\star}}\right)^{-\n}=m_a^2(T)f_a^2\, .\label{eq:qcdchi}
\end{equation} 
The high-temperature lattice QCD simulations of Ref.~\cite{Borsanyi:2016ksw} found $\chi_0^{1/4}\simeq 75.6$~MeV, $T^{\rm QCD}_{\star}\sim 150$~MeV, and $\n\simeq 7$.  Note however that other values of $\n$ have been advocated in the literature, including $\n = 8$ under the dilute instanton gas approximation~\cite{RevModPhys.53.43}, $\n=6.7$ from the interacting instanton liquid model~\cite{Wantz:2009it}, and $\n\sim 3$ from lattice calculations reported in \cite{GrillidiCortona:2015jxo}. It is likely that further studies on the lattice will be needed to refine estimates of the axion mass temperature-dependence---another reason to treat $\n$ as a free parameter. The present-day mass of the QCD axion is \cite{GrillidiCortona:2015jxo}
\begin{equation}
\label{eq:ma}
    m_a=\frac{\sqrt{\chi_{0}}}{f_a}=5.70(7) \upmu \mathrm{eV}\left(\frac{10^{12} \mathrm{GeV}}{f_{a}}\right)\, ,
\end{equation} 
and this is the value to which $m_a(T)$ in Eq.~\eqref{eq:qcdchi} settles when the temperature  drops below $T_{\star}^{\rm QCD}$.

The mass parameterisation~Eq.~\eqref{eq:qcdchi} can be readily adapted to describe more generic ALPs. These particles could arise from the spontaneous breaking of any posited global $U(1)$ symmetry {\it a priori} unrelated to the PQ solution to the strong-$CP$ problem, see, e.g., Refs.~\cite{Masso:1995tw,Masso:2002ip,Ringwald:2012hr,Ringwald:2012cu,Arvanitaki:2009fg,Cicoli:2012sz,Jaeckel:2010ni,Arias:2012az}. Some small explicit symmetry breaking---for example, in the tree-level Lagrangian, or arising from quantum effects---can give the ALP a mass, making it a pseudo-Nambu-Goldstone boson in direct analogy to the QCD axion. In keeping with the spirit of ALPs as a purely phenomenological class of particles, however, we opt in this work to be agnostic as to their physical origins, and merely take them in our simulations to be the angular mode of a complex scalar field with a mass parameterised as 
\begin{equation}\label{eq:axionmass_T}
    m_a^2(T)=m^2_a\left(\frac{T_{\star}}{T}\right)^{\n}, ~~~~~ T\geq T_f \geq T_{\star}\, .
\end{equation} 
Our model then has a generic index $\n \geq 0$, and another free parameter $m_a$ (in addition to $f_a$) that does not necessarily coincide with the relationship Eq.~\eqref{eq:ma}. We focus in this work on the range $\n \in [0,6]$, spanning between a constant-mass ALP ($\n=0$) and a model very close to the QCD axion.%
\footnote{The reason this range does not extend up to some of the predictions for the QCD axion is a matter of the computational resources needed to ensure physical behaviours at large $\n$. This issue will be discussed in Sec.~\ref{sec:units}} 

In analogy with Eq.~\eqref{eq:qcdchi}, the saturation temperature $T_{\star}$ corresponds to the  point at which the ALP mass growth stops and $m_a$ settles to its present-day value. It is in principle a model-dependent quantity in generic ALP models. However, for simplicity we shall use Eq.~\eqref{eq:axionmass_T} for the entire duration of our simulations, i.e., we assume $T_*$ to be smaller than the final simulation temperature $T_f$.  The latter generally depends on \n and corresponds to the time at which we can consider the axion field to be non-relativistic. We refer to Sec.~\ref{sec:simulationsetup} for more information on how we choose $T_f$. 

Lastly, note again that the canonical QCD axion is likely to be specified by an index in the range $\n\in[ 6.68,8]$ and has a zero-temperature mass~$m_a$ linked to the PQ symmetry breaking scale $f_a$ via Eq.~\eqref{eq:ma}.  Thus, strictly speaking all of our models are ALPs, even though we will use the term ``axion'' when discussing our simulations.

\subsection{Equations of motion}

There are two equations of motion that we evolve at different times during the simulation. Variation of the action in Eq.~\eqref{eq:action} yields the first equation of motion for the complex PQ field $\phi$,
\begin{equation}\label{eq:pqeomg}
    \ddot{\phi}+3H\dot{\phi}-\frac{\nabla^2\phi}{R^2}+\frac{\partial V}{\partial \phi}=0,
\end{equation} 
where $\cdot \equiv \partial/\partial t$ and $H\equiv \dot{R}/R$ is the Hubble expansion rate. Evolving this equation from random initial conditions leads to topological defects: cosmic strings that originate from the $V_s(|\phi|)$ contribution to the potential, and domain walls, coming from the $V_a(\theta)$ contribution, which then attach to the strings. We must solve this first equation of motion for as long as these defects are still present. However, after the network has entirely collapsed, it is convenient to integrate out the physically heavy radial mode $|\phi|$ and instead evolve the equation of motion for the axion field $a$ only, which is, 
\begin{equation}
    \ddot{a}+3H\dot{a}-\frac{\nabla^2a}{R^2}+\frac{\chi(T)}{f_a}\sin (a/f_a)=0 \, \label{eq:axiononly1}. 
\end{equation} 
The exact condition that flips the switch from tracking $\phi$ to tracking $a$ will be discussed shortly in Sec.~\ref{sec:axiononly}.

After picking up some initial value at the end of the PQ phase transition, the axion field will roll down the potential and oscillate around its minimum at $\theta = 0$. A useful reference point is the time at which the axion's zero-momentum mode begins these oscillations---put simply, when the axion becomes DM. We refer to this as the \emph{characteristic time}~$t_1$. Since this time depends on the rate of axion mass growth, our definition of $t_1$ must also be $\n$-dependent. We choose to define it starting from the relation
\begin{equation}
    c_1(\n)H_1=m_1\, ,\label{eq:c1}
\end{equation}
where $H_1 \equiv H(t_1)$, $m_1 \equiv m_a(t_1)$, and the prefactor $c_1(\n)$ is an $\mathcal{O}(1)$ correction to ensure that this notion of time is consistent for different $\n$ scenarios.

The chosen values of $c_1(\n)$ vary in the literature. In this work, we define~$t_1$ to be the time at which the solutions to the zero-momentum mode of the linearised version of Eq.~\eqref{eq:axiononly1} in two limiting regimes coincide: the solution $a=\text{const.}$ at $H\gg m_a$, and the approximate, Wentzel–Kramers–Brillouin (WKB) solution $a\propto R^{-3/2}m_a^{-1/2}$ at $H\ll m_a$. We numerically solve the linearised equation assuming radiation domination, and find that
\begin{equation}
    c_1(\n)=\frac{8}{5}\left(1+\frac{\n}{5}\right)\label{eq:cn} \, 
\end{equation}
fits the solutions reasonably well in the range $\n\in[0,6]$. A similar estimation was carried out in Ref.~\cite{Blinov:2019rhb}, which found $c_1(\n)=(2/5)(\n+4)$. This expression coincides with Eq.~\eqref{eq:cn} at $\n=0$, but differs by $\mathcal{O}(10\%)$ at $\n=6$.

\begin{figure}[t]
    \centering
    \includegraphics[width=0.99\columnwidth]{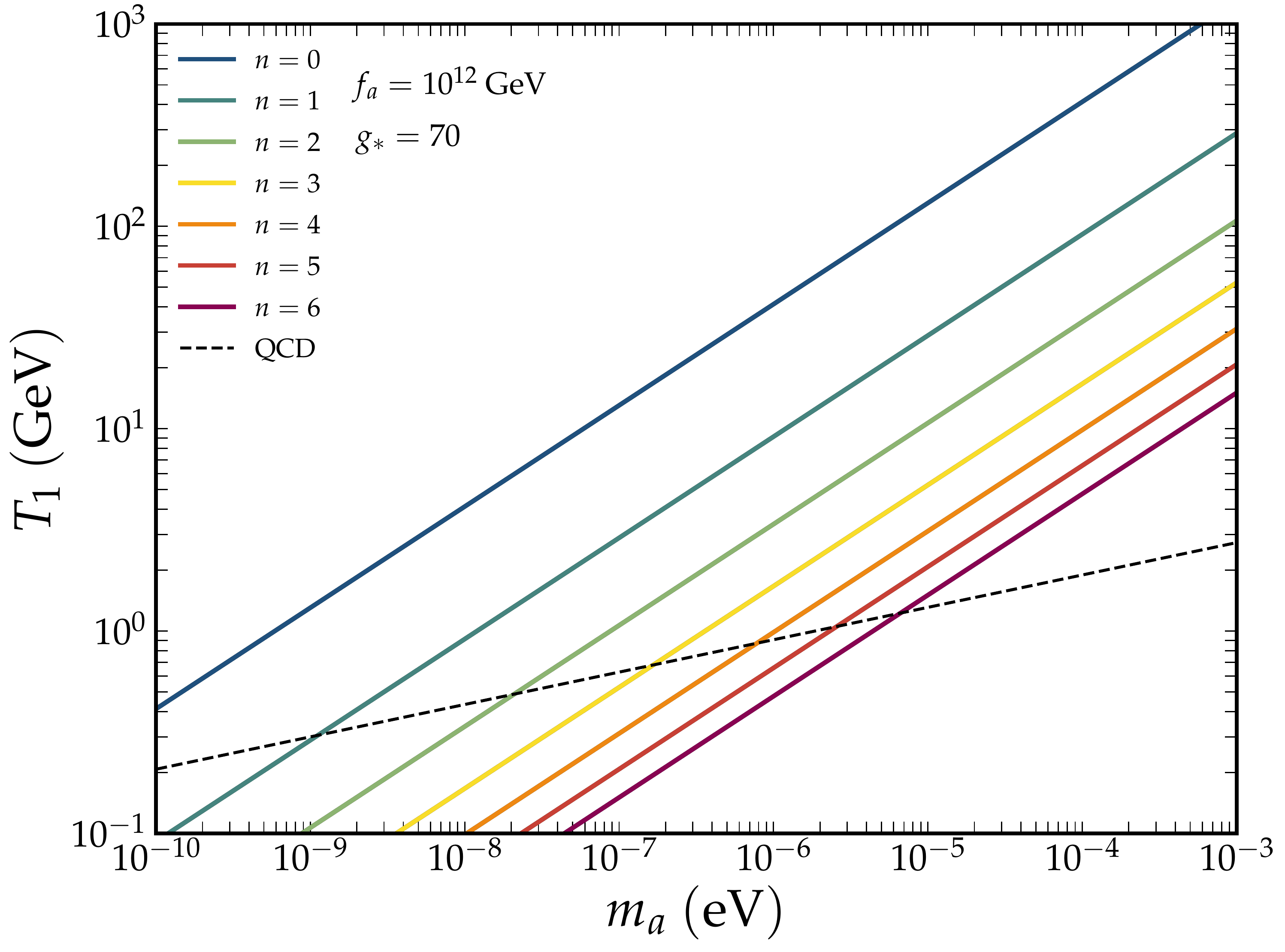}
    \caption{The characteristic temperature $T_1(\n)$, defined in Eq.~\eqref{eq:t1alps}, as a function of the present-day axion mass $m_a$.  We plot values for a range of indices $\n\in [0,8]$, with fixed $f_a=10^{12}$ GeV and $g_{\star}=70$.  The dashed line represents the QCD axion case, with numerical values adopted from Ref.~\cite{Borsanyi:2016ksw}.}
    \label{fig:t1alps}
\end{figure}

The characteristic \emph{temperature} $T_1$ that occurs at $t_1$ can now be found from Eq.~(\ref{eq:c1}) to be
\begin{equation}\label{eq:t1alps}
    T_1(\n)\simeq(m_af_a)^{1/2}\left(\frac{\sqrt{90} M_{\rm Pl}}{c_1(\n)\pi\sqrt{g_*(T_1)}f_a}\right)^{\frac{2}{\n+4}}\,, 
\end{equation} 
where $g_*(T)$ is the effective number of relativistic degrees of freedom, and $M_{\rm Pl}=2.4\times 10^{18}$~GeV is the reduced Planck mass. For the choice $\n=0$, the $f_a$ dependence cancels and we find numerically
\begin{equation}
    T_1^{\,\n=0}\simeq 40 ~\text{GeV}\left(\frac{m_a}{10^{-6}~\text{eV}}\right)^{1/2}\left(\frac{g_{*}(T_1)}{70}\right)^{-1/4}.
\end{equation} 
For the QCD axion, adopting the parameter values from Ref.~\cite{Borsanyi:2016ksw}, this yields
\begin{equation}
    T^{\rm QCD}_{1} \simeq 1.694 \,\mathrm{GeV}\left(\frac{m_{a}}{50\, \upmu \mathrm{eV}}\right)^{0.16} \,.
\end{equation} 
Figure \ref{fig:t1alps} shows typical values of $T_1$ from Eq.~\eqref{eq:t1alps} as a function of $m_a$ for various values of the index $\n \in[0,6]$ and fixed $f_a=10^{12}$~GeV and $g_*(T_1)=70$. Observe how, for ALPs, $T_1$ can be substantially larger than the reference QCD axion value, indicated by the dashed line. This relation will become important when we discuss the DM yield from our simulations in Sec.~\ref{sec:darkmatter}. 

\subsection{Units and physical scales}\label{sec:units}

We use the public code~\codo{jaxions}, originally developed in Ref.~\cite{Vaquero:2018tib}, to perform our simulations. See additional details in Appendix~\ref{sec:code}.  Following
Ref.~\cite{Vaquero:2018tib}, we work in Axion Dark Matter (ADM) units. First, we introduce the conformal time~$\eta$ via $\mathrm{d}\eta=\mathrm{d}t/R$, such that the scale factor $R$ evolves as $R \propto \eta$ during radiation domination, and $L(\eta)= \eta$ is the comoving horizon. The dimensionless simulation (code) time $\tau$ is defined as the conformal time $\eta$ normalised to its value at the characteristic time $\eta_1=L_1$, i.e.,
\begin{mdframed}[linewidth=1.2pt, roundcorner=5pt]
\vspace{-5pt}
\begin{equation}
\text{\bf Simulation time}: \quad \tau=\frac{\eta}{L_1}=\frac{R}{R_1}\,. \quad
\label{eq:tau}
\end{equation}\vspace{-0.15cm}
\end{mdframed} 
Note that the second equality here holds upon neglecting the small change in the effective degrees of freedom in the timeframe of interest~\cite{Vaquero:2018tib}. We can then express all spatial and temporal quantities in units of $L_1$ and energies in units of $H_1$. 

In ADM units it is furthermore convenient to rescale field values $\phi\to\Phi \equiv \phi\tau/f_a$ and $a\to \psi \equiv \tau a/f_a$, so that Eq.~\eqref{eq:axiononly1} can be recast in the linear regime ($\sin \theta\approx \theta$) as%
\footnote{\codo{jaxions}'s definition of the characteristic time, $\tilde{t}_1$, always corresponds to the choice of coefficient $c_1=1$ [see Eq.~\eqref{eq:c1}]. Consequently, our definition of $t_1$, which uses an $\n$-dependent $c_1(\n)$, needs to be related to the code value. We find this relation by noting that $T \propto R^{-1} \propto \eta^{-1}$ and that $\tilde{T}_1(\n)$ can be established from Eq.~\eqref{eq:t1alps} upon replacing $c_1(\n)$ with $1$. It then follows that $\alpha \equiv \tilde{\eta}_1/\eta_1=c_1(\n)^{-2/(\n+4)}$.}
\begin{equation}
    \partial^2_{\tau}\psi-\nabla^2\psi+c_1^2(\n)m^2_{\psi}\psi=0.\label{eq:c12linear}
\end{equation} 
The quantity $m_{\psi}=\tau^{(\n+2)/2}$ acts as a ``conformal'' mass. Similarly, in ADM units the full equation for the normalised complex field $\Phi$ reads \begin{equation}
    \partial^2_{\tau}\Phi-\nabla^2\Phi+\frac{\lambda}{2}\Phi(\Phi^2-\tau^2)-\tau^{\n+3}=0, 
\end{equation} 
where we have neglected the $(\partial^2_{\tau}R/R)\Phi$ term (which in any case evaluates to zero when the second equality in Eq.~\eqref{eq:tau} holds).
See Ref.~\cite{Vaquero:2018tib} for additional details.

\subsection{Discretisation and evolution}\label{sec:simulationsetup}
The simulation grids consist of a finite periodic box of comoving side length $L_c$, populated with $N^3$ homogeneously distributed points. We use $N=3072$ for all our simulations.\footnote{With the exception of one, which we will introduce in Sec.~\ref{sec:gravity}.} The physical box size, however, depends on the value of $\n$ we are simulating (for reasons that will become clear in the following section). We use $L_c = 8 L_1$ for $\n\geq4$ and $20 L_1$ for smaller $\n$, which in turn set the lattice spacing to $\Delta_x/L_1 = 2.6\times 10^{-3}$ and $6.5\times 10^{-3}$, respectively. 

We compute spatial derivatives on the lattice using a finite difference method with $n_g=2$ neighbouring points (13 stencils), accurate to $\mathcal{O}(\Delta^4_x)$. The time step $\Delta_\tau$ is dynamical, and adjusts to the maximum rate of change in the grid given by, $\omega = \sqrt{k_{\rm max}^2 + m^2}$, where the maximum momentum depends on the grid spacing as $k_{\rm max}=2\sqrt{3}/\Delta_x$. In ADM units this leads to an adaptive timestep defined by
\begin{equation}
    \Delta_{\tau}=\epsilon \left(k^2_{\rm max}+\max[m^2_{s}(\tau),m_a^2(\tau)]\tau^2\right)^{-1/2},
\end{equation} 
where the parameter $\epsilon$ can be further adjusted to improve convergence. We run our simulations using $\epsilon=1$, as convergence has been shown to be achieved for $\epsilon \lesssim 1.5$~\cite{Vaquero:2018tib}. We evolve the equations of motion in time with a four-step Runge-Kutta-Nystr\"{o}m (RKN) integrator~\cite{1992} that requires evaluations of the complex field and its derivative at three intermediate steps in order to evolve the system from a time $\tau$ to $\tau+\Delta_{\tau}$.%
\footnote{The default propagator in \codo{jaxions} is \codo{rkn4}. Other symplectic integrators include Kick-Drift-Kick (KDK) and DKD leapfrogs.} The additional iterations lead to higher-order accuracy [to $\mathcal{O}(\Delta_{\tau}^4)$] and a minimisation of the error compared to standard leapfrogs.

\subsection{Simulation duration}    
Evolving the equation of motion~(\ref{eq:pqeomg}) as is, one quickly runs into the problem that the cores of the cosmic strings, set by the saxion mass~$m_s$, remain fixed in size as the space expands around them. This means that, in the comoving coordinates used in our simulations, these cores effectively shrink over time and eventually vanish below the discretisation scale $\Delta_x$. We remedy this issue in the same way as many before us by implementing Press, Ryden and Spergel's (PRS) trick~\cite{PRS,Moore:2001px} (also referred to as the ``fat string'' trick), which is to artificially inflate the string cores over time, so that their widths---quantified by the inverse core width value, $\delta_{\rm core}^{-1}=m_s\Delta_x$---remain at a constant value in comoving coordinates. In practice, this amounts to replacing the quartic coupling in the saxion potential $V(|\phi|)$ with $\lambda \rightarrow \lambda_{\rm PRS}/\tau^2$.  A discussion of the physical implications of this procedure can be found in Ref.~\cite{Gorghetto:2018myk,Vaquero:2018tib}. At face value the trick allows us to extend the dynamical range of the simulation as much as possible, once the grid size has been chosen. Additionally, it allows us to simulate strings that are at all times thinner than \emph{physical} (not PRS scaled) strings, therefore with tension closer to its real value (see e.g. \ref{sec:initial}). Nonetheless, we recall that, despite the improvements of the PRS trick, physically extrapolated results might differ by $\mathcal{O}(1)$ corrections \cite{Gorghetto:2018myk}. Demanding that the string cores be resolved with a few lattice points immediately imposes an ${\cal O}(1)$ upper bound on $\delta_{\rm core}^{-1}$. We use values of $\delta_{\rm core}^{-1}= 1.5$ and $1$ in our simulations for $\n\geq 3$ and $n< 3$, respectively. An analysis of the impact of $\delta_{\rm core}^{-1}$ on the simulation outcome can be found in Refs.~\cite{Fleury:2015aca,Gorghetto:2018myk}.

We choose the initial time of the simulation $\tau_i$ to be the time at which the string core has a comoving (uninflated) width comparable to the comoving horizon. This condition can be written as
\begin{equation}
    \log\left(\frac{m_s}{H}\right)_{\tau_i}=0 \, ,
\end{equation}
in analogy with the time parameter $\kappa \equiv \log(m_s/H)$ used in previous studies~\cite{Gorghetto:2018ocs,Gorghetto:2020qws,Klaer:2017ond}. It then follows that the initial time and the value of $\lambda_{\rm PRS}$ (related to string tension; see Sec.~\ref{sec:strings}) depend only on the choice of the three simulation parameter values, $N,L_c/L_1,\delta_{\rm core}^{-1}$, namely,
\begin{equation}
    \tau_i=\frac{\sqrt{2}L_c/L_1}{\delta_{\rm core}^{-1}N}, ~~~~~~~~~~ \lambda_{\rm PRS}=\frac{1}{2}\left(\frac{\delta^{-1}_{\rm core}N}{L_c/L_1}\right)^2 \, .\label{eq:taui}
\end{equation} 
Apart from the fact that we use larger boxes to simulate small-$\n$ scenarios (more on this point below), these values have no explicit dependence on $\n$. We remark however that, for high-\n simulations, the value of $\lambda_{\rm PRS}$ has to be carefully chosen, taking into consideration the issue of the \emph{unphysical destruction} of domain walls. This happens if the tilting of the saxion potential, controlled by $\n$, becomes comparable to the height of the local potential maximum, determined by the saxion mass. In order to avoid such a disaster we must ensure that the condition $m^2_a/m^2_s<1/40$ is satisfied by our simulations when topological defects are present~\cite{Fleury:2015aca}.  For $\n=6$, this means $N=3072$ at minimum, and is also the main reason for our choice of the grid resolution.

Naively, the finite volume of the simulation box would dictate that we can simulate only up to $\tau = (L_c/L_1)/2$, i.e., the time it takes for a wave emitted from a point, travelling at the speed of light, to bump into and interfere with itself. However, because the axion and saxion fields are both massive, the simulation time is ultimately not limited by the speed of light, but by the group velocity of the fastest modes in the box and their associated free-streaming scales. For a given mode $k$, the free-streaming scale~$\lambda_{\mathrm{fs}}$ in ADM units can be calculated as%
\footnote{This formula holds only up until the time the mass growth saturates, $\tau_{\star}$. The contribution to the integral at times $\tau>\tau_{\star}$ is the same, but with $\n\to 0$ and $\tau_{\star}$ as the lower integration bound.}
\begin{equation}
    \lambda_{\mathrm{fs}}(k,\tau;\n)/L_1=\int_0^{\tau}{\rm d} \tau^\prime \,\left(1+\frac{{\tau^\prime}^{\n+2}}{(kL_1)^2}\right)^{-1/2}.\label{eq:fslen}
\end{equation} 
Therefore, if we wish to simulate up to a particular final time~$\tau_f$, we can adjust the simulation box size and resolution such that the condition $L_c/L_1\gtrsim 2\lambda_{\mathrm{fs}}(k,\tau_f)$ is satisfied by the bulk of the available modes in the box.

\begin{figure}[t]
    \centering
    \includegraphics[width=0.99\columnwidth]{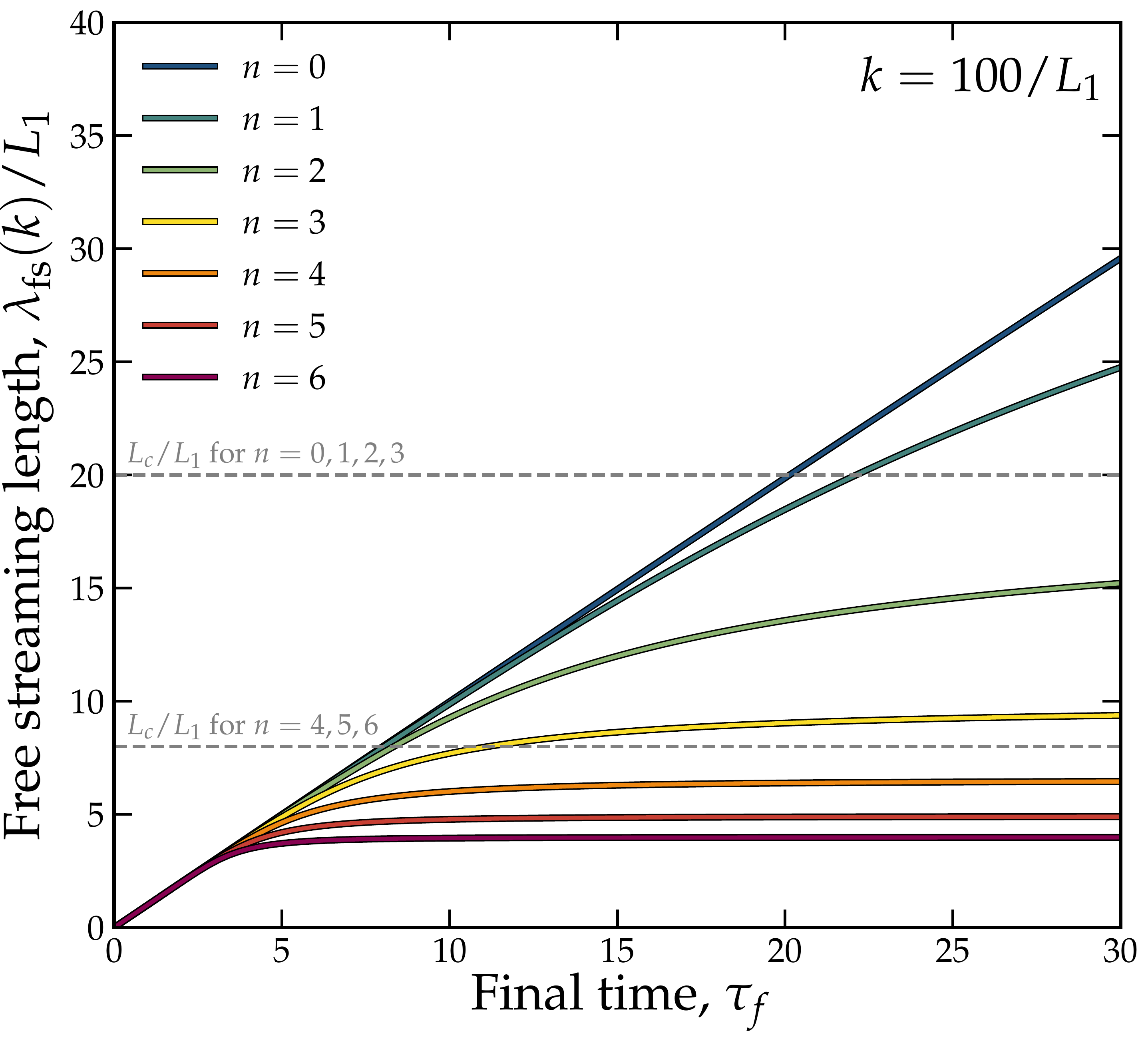}
    \caption{Free-streaming length $\lambda_{\rm fs}$ in units of $L_1$, as defined in Eq.~\eqref{eq:fslen}, for a mode with momentum $k = 100/L_1$. We show the free-streaming length as a function of the final simulation time $\tau_f$ for our range of indices~$\n$.}
    \label{fig:cn_fs}
\end{figure}

Figure~\ref{fig:cn_fs} shows an example of $\lambda_{\rm fs}$ for the mode $k=100/L_1$ as a function \n and the final time $\tau_f$. Clearly, there is no substantial free-streaming at this $k$ mode for the larger values of $\n$.  Furthermore, it has been argued in Ref.~\cite{Vaquero:2018tib} that finite volume interference effects from momenta larger than $k\sim 100/L_1$ do not affect substantially the output power spectrum, and are suppressed in the \emph{adiabatic} (i.e., WKB) approximation that we implement in Sec. \ref{sec:adiabatic}. Therefore, for the $\n\geq4$ simulations, a box length of $L_c = 8 L_1$ suffices. On the other hand, the free-streaming lengths for low values of $\n$ are substantial, and larger boxes must be employed to keep the simulation running for a longer time free of interference effects.  Since we anticipate that, for $\n\lesssim 3$, a full study of the axion field dynamics after the network collapse necessitates that we simulate to $\tau\gtrsim 5$, we choose a simulation box size of $L_c=20 L_1$ for these simulations. This gives a Nyquist momentum of $k_{\rm Ny}=\pi/\Delta_x\simeq 483/L_1$, close to the example $kL_1=100$ shown in Fig.~\ref{fig:cn_fs}.

Having fixed the box size and resolution, we can now choose the final simulation time $\tau_f$ by ensuring that most of the modes in the box have transitioned to the non-relativistic regime in time. Because for low-\n runs the Nyquist frequencies will undergo this transition at times we cannot simulate, we set our minimum~$\tau_f$ by way of the condition $k_{\rm Ny}/y=m_{\psi}(\min{\tau_f})$, where $y\sim 4$~\cite{Vaquero:2018tib}; all momenta smaller than $k_{\rm Ny}/y$ can be considered to be non-relativistic at $\tau=\min{\tau_f}$. This lower bound can be equivalently expressed as
\begin{equation}
    \min{\tau}_f=\tau_{\mathrm{Ny}/y}=\left(\frac{\pi}{y\Delta_x}\right)^{\frac{2}{\n+2}} \,.\label{eq:mintf}
\end{equation} 
Thus, the value of the final simulation time decreases with $\n$ according to this criterion. 

Another consideration in setting the final simulation time relates to a class of structure that we have not yet introduced, but will come to dominate the power spectrum at the latest times. These are the \emph{axitons}, which will be discussed in detail in Sec.~\ref{sec:axitons}. For now it suffices to say that axitons are very small and dense lumps that require $\Delta_x\ll \pi/m_{\psi}$ to resolve. Since $m_\psi$ increases with time, there will be a moment beyond which our choice of $\Delta_x$ is no longer sufficient to resolve the axitons. This effectively sets 
\begin{equation}
    \max\tau_f\sim \Delta_x^{-\frac{2}{\n+2}}\label{eq:maxtf}
\end{equation} 
as a \emph{maximum} that our final simulation time must not exceed.

It is important to recognise that, by the criteria~\eqref{eq:mintf} and \eqref{eq:maxtf},  the situation $\min{\tau_f} > \max{\tau_f}$ could arise if $y<\pi$, in which case satisfying one criterion must necessarily violate the other. Conversely, for the choice of $y\gtrsim 4$, the minimum and maximum times are quite close to one another in the large-$\n$ cases.
It is therefore a nontrivial task to choose the final simulation times over our range of \n while satisfying both Eqs.~\eqref{eq:mintf} and \eqref{eq:maxtf} simultaneously. We find that $\tau_f\in [5,5.5]$ for $n\geq 4$ and $\tau_f\in [8,12]$ for $\n<4$ are suitable. 

\subsection{Initial conditions} \label{sec:initial}

\begin{table}[t]\centering
\ra{1.3}
\begin{tabularx}{0.45\textwidth}{Y|YYYYY}
\hline \hline
$\n$ & $L_c/L_1$ & $\delta^{-1}_{\rm core}$ &  $\tau_2$    &  $\tau_f$   &  $\tau_{\rm ad}$  \\ \hline
 0   &       20          &         1.0                 &     2.4      &     12.0    &      50.0         \\
 1   &       20          &           1.0              &      2       &     12.0    &      30.0       \\
 2   &       20          &          1.0                &     1.8      &     10.0    &      20.0         \\
 3   &       20          &          1.5                &     1.65     &     8.0     &      12.0          \\
 4   &        8          &            1.5                &     1.55     &     5.0     &      10.0        \\
 5   &        8          &           1.5                 &     1.48     &     5.5     &      8.0       \\
 6   &        8          &          1.5                  &     1.42     &     5.0     &      8.0      \\
 \hline \hline
\end{tabularx}
\caption{Summary of the simulation parameters for each of our simulations covering values of $\n$ from 0 to 6. The parameters are, from left to right, the box size $L_c$, the inverse string core width $\delta^{-1}_{\rm core}$, the time at which the strings and domain walls have equal densities~$\tau_2$, the final time of the full evolution $\tau_f$, and the true final time  to which we extend  via the adiabatic approximation $\tau_{\rm ad}$.}\label{tab:params}
\end{table}
\begin{figure*}[t]
    \centering
    \includegraphics[width=0.98\textwidth]{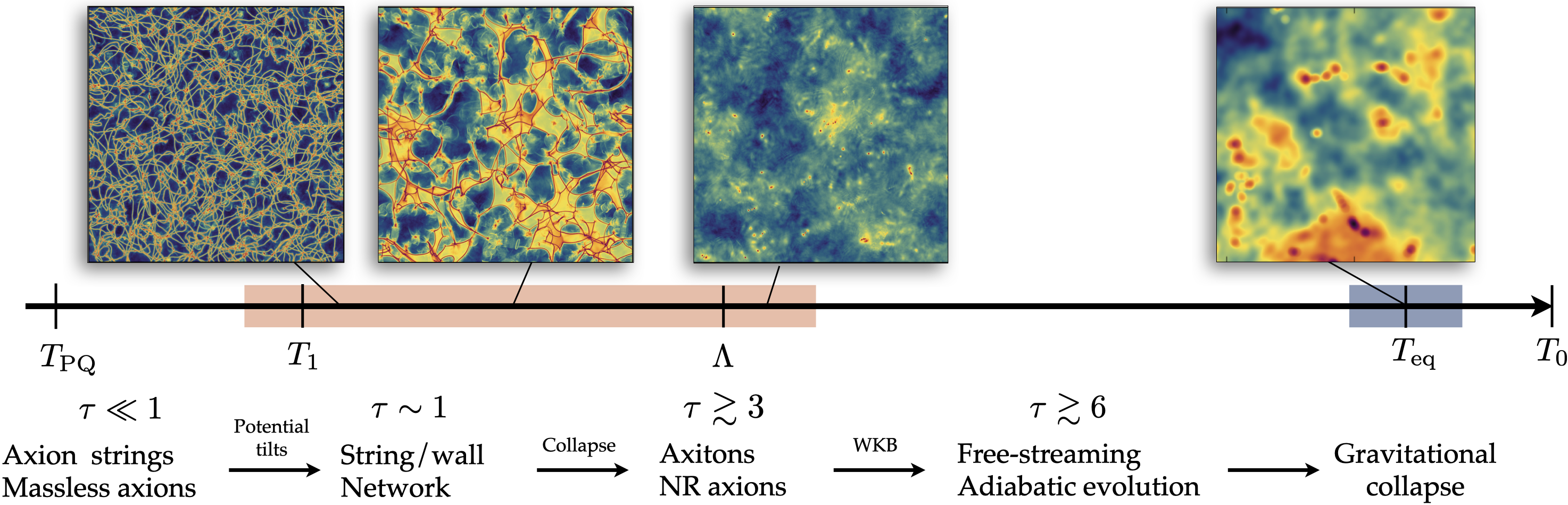}
    \caption{General timeline for the evolution of the axion field after PQ symmetry breaking, through the QCD phase transition, until the eventual gravitational collapse of inhomogeneities into miniclusters. The era studied here is highlighted in orange, whereas the era requiring devoted $N$-body simulations is highlighted in blue. We discuss each of these stages in detail in Sec.~\ref{sec:dynamics}.}
    \label{fig:timeline}
\end{figure*}

The final input to the simulation to be discussed are the initial conditions. Properly initialising the grid for the complex field $\phi$ would require that we include finite-temperature corrections that dominate the full potential at $T\gg f_a$ and then slowly disappear after the PQ phase transition at $T\sim f_a$; see for example the approach taken in Refs.~\cite{Hiramatsu:2012gg,Buschmann:2019icd}.  However, because our choice of the initial time [Eq.~\eqref{eq:taui}] corresponds to $T_i\ll f_a$, these temperature corrections can be completely neglected.  

\codo{jaxions} has a class to generate initial conditions in several ways. Our simulations are initialised in momentum space, where the Fourier field $\tilde{\phi}(k)=|\tilde{\phi}|e^{i \theta}$ has a randomly chosen phase $\theta$ in the interval $[-\pi,\pi]$, $\langle\tilde{\phi}\rangle =0$, and $|\tilde{\phi}|$ is drawn from a half-Gaussian that peaks at $|\tilde{\phi}|=0$ and has a variance parameter 
\begin{equation}
    \sigma^2_{\vert\tilde{\phi}(k)\vert}=\exp(-k^2/k^2_{\rm cr})\,.
\end{equation} In other words, $\tilde{\phi}(k)$ has a flat spectrum on large scales but is cut off at a characteristic length scale corresponding to the average inter-string separation $\sim \pi/k_{\rm cr}$. The latter is closely related to the string density parameter $\xi$ (to be defined in Sec.~\ref{sec:strings}), whose value follows the attractor solution~\cite{Gorghetto:2018myk,Gorghetto:2020qws}
\begin{equation}
     \xi(\kappa)=c_{-2}\kappa^{-2}+c_{-1}\kappa^{-1}+c_0+c_1\kappa,\label{eq:xifit}
 \end{equation} 
where $c_i$ are fitting coefficients and $\kappa\equiv \log(m_s/H)$ is the string tension.

Observe that our choice of the initial conditions does not strictly mimic the field configuration one would expect at values of the string tension we are able to simulate with current computational resources, i.e., $\kappa\lesssim 8$.  However, because the statistics of the initial power spectrum are consistent with that of the attractor solution, once the simulation begins the resulting string density quickly approaches the attractor solution, independently of the box size, resolution, or other simulation parameters. Extrapolating to physical values of the string tension---e.g., $\kappa\sim 70$ for the QCD axion, but in general for ALPs this could be substantially different---Refs.~\cite{Hindmarsh:2019csc,Hindmarsh:2021vih,Hindmarsh:2021zkt} find $\xi\simeq 1.2$ for the string density parameter, a value very close to what we observe in our simulations at $\tau\lesssim 1$. We remark however that there is no numerical evidence of the value of $\xi$ at large values of the string tension (as for the spectral index of the axion emission spectrum) and the value of $\xi$ could be much larger, e.g., $\mathcal{O}(15)$, as supported by the arguments of Refs.~\cite{Gorghetto:2018myk,Gorghetto:2020qws}. In the following we do not focus on this important issue, as we compare different $\n$ scenarios, assuming the string tension and density that can be achieved with our current computational power. 

\subsection{Comparison with previous simulations}
Before we move to our results, we will briefly comment here on the differences between our simulation setup and those found in prior literature. The major distinguishing factor of our simulations is that we consider a general parameterisation for the axion mass, as opposed to single power law, however there are several previous works that share our general goals but differ in their implementation. As we have already mentioned, our simulations can be considered a continuation of Ref.~\cite{Vaquero:2018tib} but for a range of $\n$ hence we need not spend time discussing the difference between this study. However Ref.~\cite{Hiramatsu:2012gg,Fleury:2015aca,Klaer:2017ond,Buschmann:2019icd} also studied the evolution of the axion field through the string scaling and QCD eras, but with a few important differences in setup. Refs.~\cite{Hiramatsu:2012gg,Buschmann:2019icd} do not exploit the PRS trick, favouring a fixed string width, however this does not eliminate the issue of the chosen value of $\lambda$ is still unphysically small. Finally, the simulations presented in Refs.~\cite{Kawasaki:2018bzv,Gorghetto:2018myk,Gorghetto:2020qws,Hindmarsh:2019csc,Buschmann:2021sdq} are not directly comparable to ours as they consider only the PQ string scaling era with a zero axion mass and extrapolate based on the string emission spectrum.

\section{Dynamics of the field}\label{sec:dynamics}

Now that we have discussed all of the key ingredients to the simulations we can finally begin to discuss our results. For reference we have listed all of the key parameters discussed so far in Table~\ref{tab:params}. 

In this section we will outline both qualitatively and quantitatively the major epochs that the PQ field $\phi$ undergoes in chronological order. A rough timeline is illustrated in Fig.~\ref{fig:timeline} (the values of $\tau$ are not precise). In particular, we will take note of the differences in the way this evolution proceeds when we change the value of $\n$ in our model. Recall that the case $\n = 0$ can be thought of as a generic temperature-independent ALP, whereas the value $\n = 6$ corresponds to a rate of mass growth very similar to the QCD axion. A complete picture of the evolution of the axion field over our range of $\n$ is given in  Appendix~\ref{sec:extra_viz}.  

\subsection[Axion strings]{\texorpdfstring{$\tau\ll1$}: axion strings}\label{sec:strings}

The dynamics of the PQ field at early times $\tau\ll 1$ can be described entirely by the equation of motion~\eqref{eq:pqeomg} under the saxion potential $V_s(|\phi|)$ alone. 
After the spontaneous breaking of the PQ symmetry at $T \sim f_a$, a network of global axion strings appears as described by the Kibble mechanism~\cite{Kibble:1976sj}. These strings are topological defects on spatial lines around which the phase of the PQ field, $\theta= a f_a$, wraps $2\pi$ in field-space.  Subsequent evolution sees these linear defects decay into relativistic modes of the angular part of the PQ field, i.e.,~axions, as well as a small amount ($\sim 15\%$) in the radial saxion mode~\cite{Gorghetto:2018myk}. At this time, the axion mass $m_a(T\gg T_1)$ is small enough so as not to affect the strings' evolution and can therefore be neglected from the equation of motion.

To describe the global properties of the string network, it is conventional to define the dimensionless string density parameter $\xi = \ell(t) t^2/\mathcal{V}$, that represents the number of strings of physical length $t\sim H^{-1}$ per causal volume $\sim t^3$, expressed in terms of their total length $\ell(t)$ over a physical simulation volume $\mathcal{V}=(L_cR)^3$. In ADM units this is equivalently
\begin{equation}
    \xi(\tau)=\frac{n_{\rm plaqs}\Delta_x\tau^2}{6(L_c/L_1)^3}\,,\label{eq:xicode}
\end{equation}
where $n_{\rm plaqs}$ denotes the number of plaquettes pierced by a string. We identify the string locations using the algorithm of Ref.~\cite{Hiramatsu:2010yu,Fleury:2015aca}. As explained in detail in Ref.~\cite{Fleury:2015aca}, the numerical factor in Eq.\eqref{eq:xicode} divides $n_{\rm plaqs}$ by the angle-averaged number of plaquettes per unit length of string, and as such Eq.~\eqref{eq:xicode} represents a statistical estimate of $\xi$.

The network is expected to follow a scaling solution in which $\xi$ is a constant $\mathcal{O}(1)$ number~\cite{Yamaguchi:1998gx,Yamaguchi:1999yp}. However, as observed in many recent simulations~\cite{Fleury:2015aca,Gorghetto:2018myk,Vaquero:2018tib,Buschmann:2019icd}, the $\xi$ parameter exhibits a small logarithmic increase over time, following the same factor $\kappa=\log(m_s/H)$ that appears in the string tension and string-axion coupling. As  discussed already in the context of our initial conditions, Ref.~\cite{Gorghetto:2018myk} showed that the network of axion strings exhibits an attractor solution given by the fitting function~\eqref{eq:xifit}. For $\kappa\gtrsim 4$, this can be approximated as $\xi(t)=c_1\kappa+c_0$, with $\mathcal{O}(1)$ coefficients $c_{0,1}$. Similarly, the energy density in strings at this stage of the evolution is expected to scale with cosmic time $t$ as \cite{Gorghetto:2018ocs}
\begin{equation}
    \rho_s(t)=\frac{\xi\mu_{\rm{eff}}}{t^2}\simeq \frac{\xi\pi f^2_a\kappa}{t^2},\label{eq:stdensity}
\end{equation} 
where $\mu_{\rm{eff}}\simeq\pi f^2_a\kappa$ is the effective string tension. 

We observe the value of $\xi$ to scale linearly with $\kappa$, in agreement with previous works~\cite{Fleury:2015aca,Klaer:2017ond,Gorghetto:2018myk,Vaquero:2018tib,Buschmann:2019icd,Gorghetto:2020qws}. This holds up until $\tau\gtrsim 1$, where the explicit breaking of the PQ symmetry due to the presence of the axion potential $V_a(\theta)$ causes deviations from the scaling regime, as we will now discuss.

\subsection[Strings and domain walls]{\texorpdfstring{$\tau\gtrsim 1$}: strings and domain walls}

\begin{figure}[t]
    \centering
    \includegraphics[width=0.99\columnwidth]{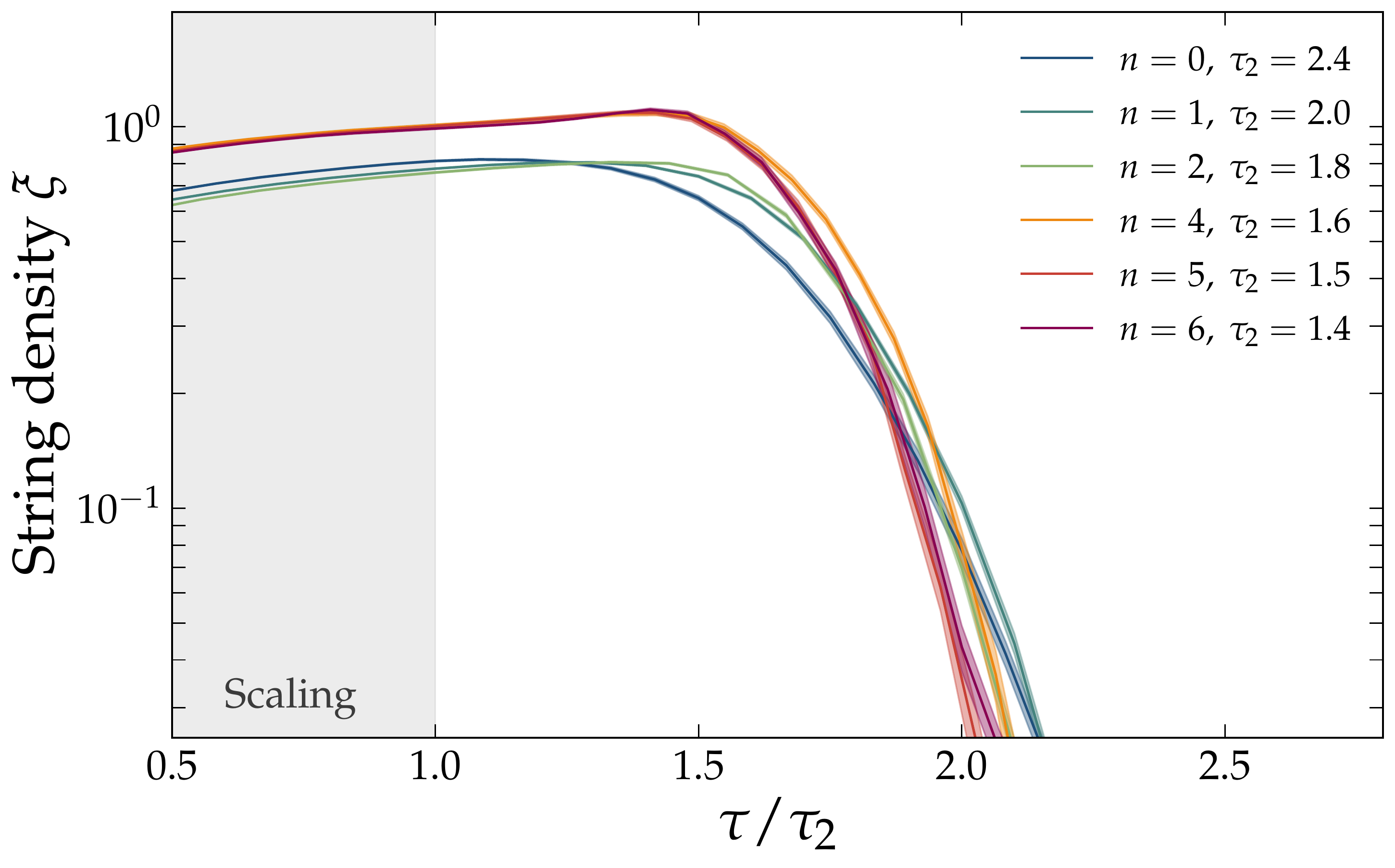}
    \caption{Scaling of the string density $\xi(\tau)$ during the early stages of the simulations. We normalise the simulation time $\tau$ here by $\tau_2$ [see Eq.~\eqref{eq:tau2}], which is the time at which the energy densities in the strings and domain walls are equal. At $\tau/\tau_2\lesssim 1$ the strings dominate the energy density, whereas at $\tau/\tau_2\gtrsim 1$ the domain walls begin to dominate, causing the network to collapse. The value of $\tau_2$ depends on $\n$, so that displaying the scaling of $\xi$ as a function of this rescaled time allows us to confirm our expectation from geometric arguments that the network should collapse around $\tau = 2\tau_2$. The band around each curve is the standard error collected from ten simulations per  choice of~\n.} 
    \label{fig:scaling}
\end{figure}

Around the characteristic time $\tau\sim 1$, the axion potential starts to influence the evolution of the field. Domain walls---defined as surfaces on which $\theta\sim \pi$---form between strings of different chiralities. The surface tension of the walls acts to pull the strings together, ultimately causing the collapse of the entire string-wall network. The faster the axion mass grows (i.e., the larger $\n$ is), the faster the topological defects can be expected to disappear. 

In general, stable domain walls are disastrous for cosmology as they would grow to exceed the allowed present-day energy budget. Fortunately, however, in the post-inflationary scenario the axion string-wall network is unstable if the domain-wall number is $N_{\mathrm{DW}}=1$. This is the case for axion models where only one vacuum is created by the explicit breaking, i.e., strings attach to the same domain wall.\footnote{Another way in which the walls can be destabilised is if the potential contains a \emph{bias} term---either induced by gravitational effects or by higher-dimensional operators---that effectively makes one of the $N_{\textrm{DW}}$ degenerate vacua the true vacuum, see, e.g., Refs.~\cite{Barr:1982uj,Lazarides:1982tw,Dvali:1994wv,Chang:1998bq,Barr:2014vva,Reig:2019vqh,Caputo:2019wsd,Gelmini:1988sf,Gelmini:2021yzu,Chen:2021wcf,Chen:2021hfq}.} We are assuming $N_{\mathrm{DW}}=1$, which is the case for the popular hadronic class of QCD axion model~\cite{Kim:1979if,Shifman:1979if} and is not an unreasonable assumption for a generic ALP as well. 

Like its string counterpart $\xi$, the cosmological wall density is observed to be an $\mathcal{O}(1)$ number in our simulations. However, the energy density of the domain walls scales differently,
\begin{equation}
    \rho_w(t)\simeq\frac{\sigma(t)}{t}, \;\;\;\;\;\;\ \sigma(t)=8m_a(t)f^2_a,\label{eq:walldensity}
\end{equation}
where $\sigma$ is the wall's surface tension, i.e., energy per unit area. Compared with the string energy density $\rho_s \propto t^{-2}$ given in Eq.~\eqref{eq:stdensity}, we see that the wall energy density $\rho_w \propto t^{-1}$ must eventually dominate. We can therefore define another useful timescale, $t_2$, as the time at which the wall and string energy densities are equal, i.e.,  $\rho_w(t_2)=\rho_s(t_2)$. In ADM units, this timescale is equivalently
\begin{mdframed}[linewidth=1.2pt, roundcorner=5pt]
\vspace{-5pt}
\begin{equation}\label{eq:tau2}
    \tau_2=\left(\frac{\pi\kappa}{4}\right)^{\frac{2}{\n+4}} \,.
\end{equation}\vspace{-0.15cm}
\end{mdframed} 
As the wall tension begins to dominate, the remaining lifetime of the strings is limited by the time it takes the walls to pull them together by a distance $r_H(\tau_2)$. We therefore expect the string-wall network to have collapsed by $\tau = 2 \tau_2\in [5,2.8]$, where the range corresponds to our considered range of indices $\n\in[0,6]$ (see Table~\ref{tab:params}). Equation~\eqref{eq:tau2} suggests that models with slower mass growth will have a delayed network collapse compared to models with a faster mass growth. Figure~\ref{fig:scaling} confirms this expectation, where we see that the string network disappears around $\tau/\tau_2\sim 2$ in all cases. Notice also how in the $\n=0$ case the collapse begins slightly before $\tau/\tau_2\sim 1.5$, resulting in a smoother and longer-lasting network destruction compared with the abrupt turnover observed in the $\n=5$ and 6 cases. 

\subsection[Axion-only simulation]{\texorpdfstring{$\tau\gtrsim 3$}: axion-only simulation}\label{sec:axiononly}

After the strings and walls have collapsed, the saxion has fulfilled its role and the field can be suitably described by the axion oscillating around its minimum at $\theta=0$. At this point we only need to keep track of the angular degree of freedom of~$\phi$, as done for instance in Refs.~\cite{Vaquero:2018tib,Buschmann:2019icd}. In practice, the switch occurs when no plaquette is tagged as containing topological defects. After the switch, we solve instead the equation of motion Eq.~\eqref{eq:axiononly1}, with a potential containing essentially a mass term and an attractive self-interaction.  

In order to study large gradients, $\nabla\theta\sim\mathcal{O}(\pi/\Delta_x)$, we evolve the equation on the unbounded domain $\theta\in(-\infty,\infty)$. So after the switch from $\phi$ to $\theta$ we add values of $2\pi$, until a continuous $\theta$ field is obtained across the grid.%
\footnote{In \codo{jaxions}, axion-only evolution can also be performed in periodic mode, where $\theta\in[-\pi,\pi]$.}  

Since the axion mass is still increasing with the time at this stage of the evolution, modes will gradually transition from relativistic to non-relativistic. Once the bulk  of the modes has undergone this transition, we can study the distribution of overdensities as one would for normal cold DM. We can estimate the axion energy density $\rho_a$ at these times by noting that, 
once $\phi$ develops a vev $\langle\phi\rangle$, the massive mode $\vert\phi\vert$ can be integrated out, so that the energy left in the $\phi$ field corresponds to
\begin{equation}
    \rho_a=\frac{1}{2}\dot{a}^2+\frac{1}{2 R^2}(\nabla a)^2+\chi(1-\cos a/f_a)\,.\label{eq:endensity}
\end{equation} 
The axion \emph{density contrast} can then be defined as
\begin{equation}
    \delta_a(x)=\frac{\rho_a}{\langle \rho_a\rangle}-1,
\end{equation}
where $\langle \rho_a\rangle$ is the spatial average of the energy density stored in the axion field.

\begin{figure}[t]
    \centering
    \includegraphics[width=0.99\columnwidth]{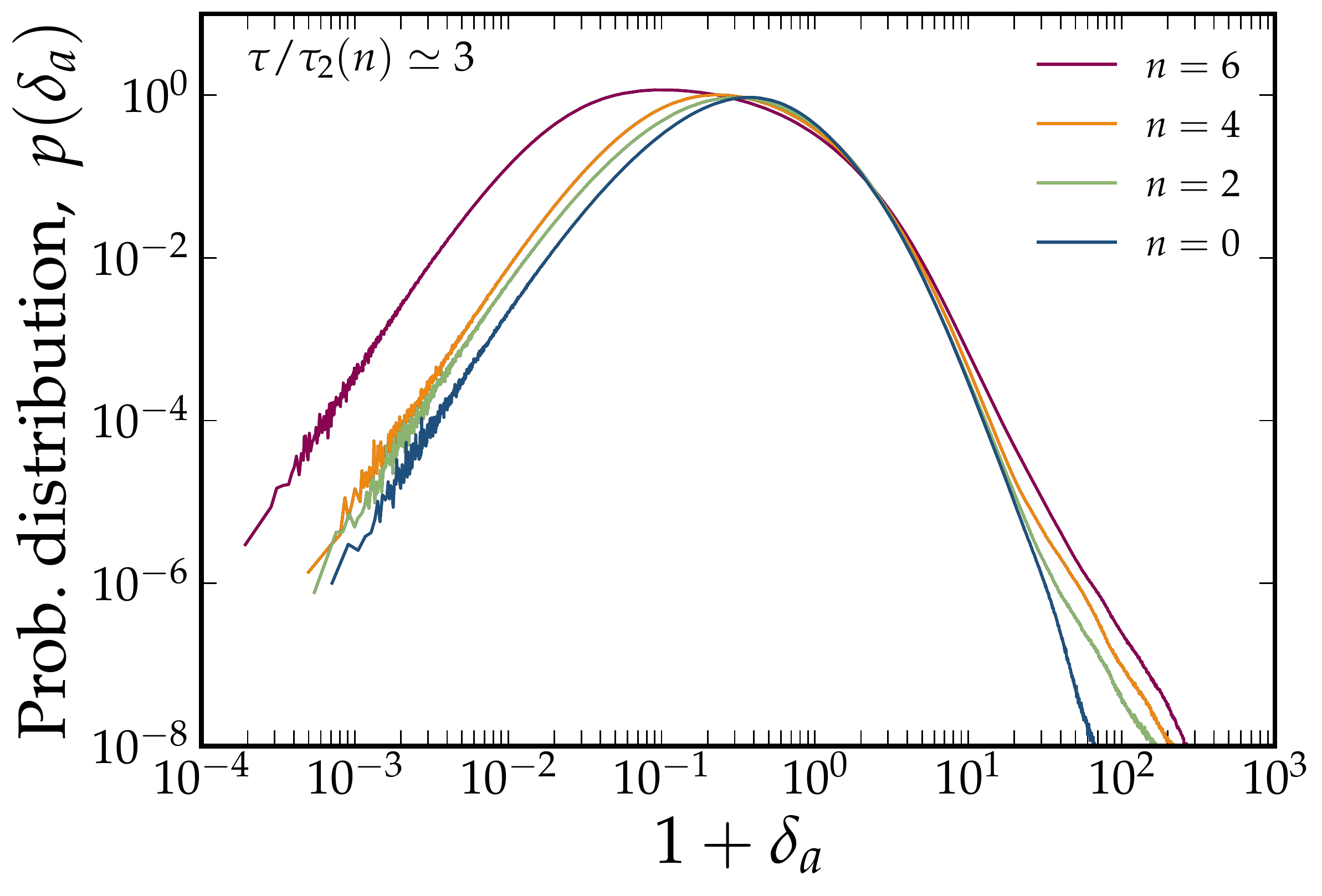}
    \caption{Distribution of the axion density contrast $\delta_a$ at the simulation time $\tau/\tau_2(\n)\simeq 3$ for four of the seven values of \n that we simulate. The cases $\n = 1$, 3, 5 have been removed for legibility. We observe that the temperature-independent ALP case ($\n = 0$) peaks at slightly larger values of $\delta_a$. However, the distribution is quickly suppressed away at values of $\delta_a$ beyond the peak. In general, we observe the trend that faster growing axion masses lead to significantly more instances of high density fluctuations, i.e., $1+\delta_a \gtrsim 100$, due to the appearance of axitons in these cases.}
    \label{fig:delta}
\end{figure}

Figure~\ref{fig:delta} shows the probability distribution $p(\delta_a)$ of the density contrasts evaluated on the grid sites of one sample simulation from each of our $\n=0,2,4,6$ runs at the normalised simulation time $\tau/\tau_2\simeq 3$.  We shall defer  the discussion of the \emph{spectrum} of these energy density fluctuations to Sec.~\ref{sec:specshape}.  But we note here that the $\n=0$ case clearly supports far fewer very high density regions (i.e.,  where $1+\delta_a \gtrsim 100$) than higher-\n cases, and that increasing \n generally leads to larger overdensities.  This trend can be understood as follows.

\begin{figure*}[t]
    \centering
    \includegraphics[width=0.99\textwidth]{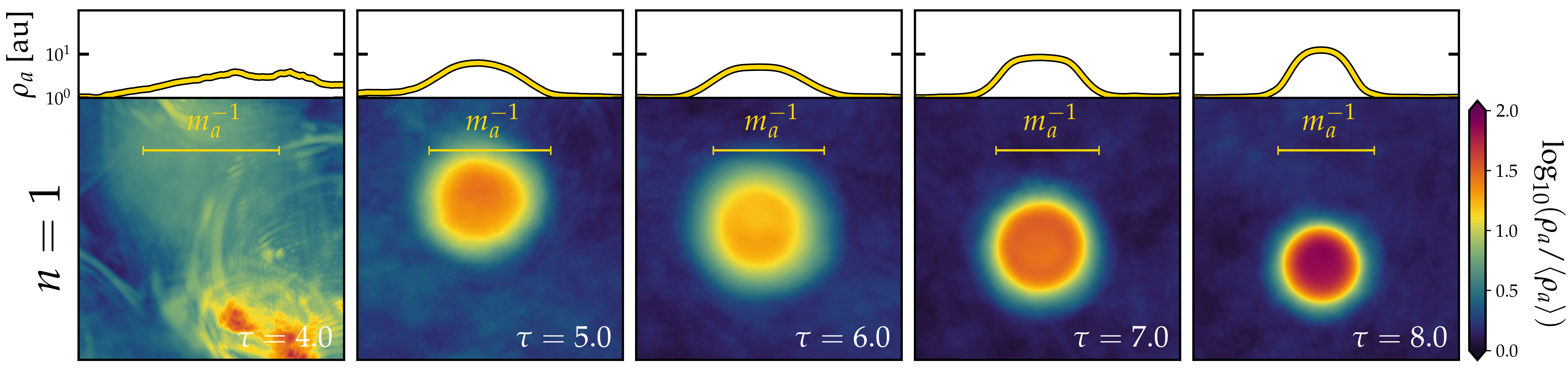}
    \includegraphics[width=0.99\textwidth]{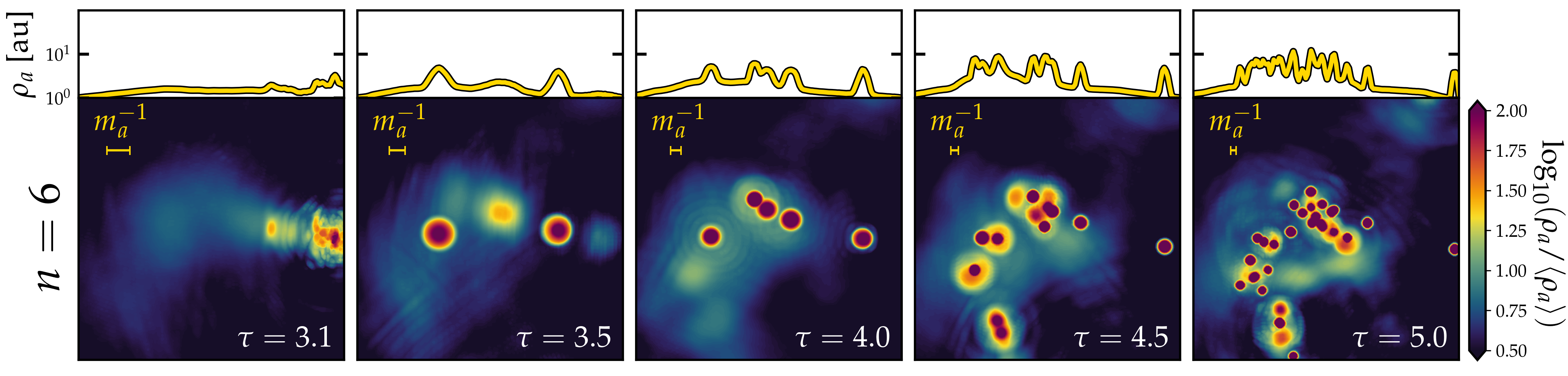}
    \caption{Evolution of a $150\Delta_x \times 150\Delta_x$ region of the projected axion density as a function of time ($\tau$ running from left to right). The upper panels show a zoom-in of a solitary axiton in an $\n=1$ simulation, whereas the lower panels show the formation of a large cluster of axitons in an $\n=6$ simulation. Above each 2D density projection, we also show a 1D projection of the density to further highlight the profiles of the axitons. In particular, in the upper panels where the axiton is large relative to $\Delta_x$, we see that its shape varies with time as it shrinks whilst also radiating axions radially.}
    \label{fig:axitonprofile_n12}
\end{figure*}

Around this time, the field in the faster growing axion mass models (i.e., large $\n$)  begins to be dominated by a new kind of structure with very large overdensities. These structures are the axitons that we hinted at earlier: quasi-stable oscillating configurations of the axion field~\cite{Kolb:1993hw,Visinelli:2017ooc}. They are related to a more generic class of solutions to the three-dimensional Sine-Gordon equation with an increasing mass, also known as oscillons or pseudo-breathers. Axitons have sizes comparable to the axion's Compton wavelength $\sim 1/m_a$. As they are supported by the attractive self-interaction, they will persist in the simulation, shrink, radiate axions, and seed other axitons for as long as the axion mass continues to grow.  Figure~\ref{fig:axitonprofile_n12} shows several snapshots of a small region of the projected density in two simulations, each containing a sample of these axitons. The radiation of axions from the axitons is particularly clear in the $ \tau =4.0$ snapshot of the $\n=6$ case (bottom row, third panel from left), which is followed at $\tau=4.5$ and $\tau=5.0$ by many more axitons forming around them. In contrast, axitons in the slow mass growth scenarios are generally isolated from each other and less plentiful overall (see upper panels of Fig.~\ref{fig:axitonprofile_n12} for the case of $\n=1$), in agreement with our density fluctuation count in Fig.~\ref{fig:delta}.  

\subsection[Adiabatic evolution]{\texorpdfstring{$\tau\gtrsim6$}: adiabatic evolution}\label{sec:adiabatic}

Once most of the axion modes have become non-relativistic and $\theta\ll 1$, we can continue the evolution analytically by linearising the equation of motion,  i.e., approximating $\sin\theta\approx\theta$ in Eq.~\eqref{eq:axiononly1}.
 Using the conformally rescaled field $\psi=\tau\theta$, each of its Fourier modes $\tilde{\psi}_k$ evolves independently by the equation of motion~\cite{Vaquero:2018tib},
\begin{equation}
    \partial^2_{\tau} \tilde{\psi}_k+w^2_k\tilde{\psi}_k=0, ~~~~~~~~~~w^2_k=k^2L_1^2+c_1^2\tau^{\n+2}\,.
\end{equation}
In the so-called \emph{WKB approximation} or, equivalently, the adiabatic limit $\partial_{\tau}w_k/w_k^2\ll 1$ (which is satisfied in all our runs at $\tau \gtrsim 6$), the solution can be written as
\begin{equation}
    \tilde{\psi}_k(\tau)=\sum_{\pm}c_{\pm}\sqrt{\frac{w_k(\tau_0)}{w_k(\tau)}}e^{\pm i\varphi(\tau)},  ~~~ \varphi(\tau)=\int_{\tau_0}^{\tau}w_k(\tau')d\tau',\label{eq:wkb}
\end{equation}
where the constants $c_{\pm}$ are fixed by initial conditions. Identifying $\tau_0$ with the final simulation time $\tau_f$ (see Table~\ref{tab:params}), the phase integral has an analytical solution,
\begin{equation}
    \varphi(\tau)=\left[\frac{2w_k\tau}{4+\n}\left(1+\frac{2+\n}{2}\frac{k^2}{w_k^2}{}_{2}F_1(1/2,1,a_1,a_2)\right)\right], \label{eq:phase}
\end{equation}
where ${}_{2}F_1$ is the hypergeometric function, with arguments \begin{equation}a_1=1+\frac{1}{2+\n},~~~~~~a_2=m^2_{\psi}/w^2_k.
\end{equation}
We use the WKB solution~\eqref{eq:wkb} to extend the simulation outcome to $\tau_{\rm ad}\in[5,50]$, where the larger value corresponds to the $\n=0$ case.

Since the WKB approximation switches off the axion self-interactions, it removes the axitons (see Sec.~\ref{sec:axitons}) that usually survive past $\tau_f$, as well as power in very high $k$ modes \cite{Vaquero:2018tib}. The disappearance of axitons is expected in the QCD axion case once the axion mass---or more precisely topological susceptibility---approaches the zero-$T$ values. This was confirmed by the simulations of Ref.~\cite{Buschmann:2019icd}, where the authors explicitly implemented a cutoff in the axion mass growth within the simulation. For a general choice of $\n$, we interpret the disappearance of axitons from our WKB result to be analogous to the QCD case.  Then, in the absence of self-interaction, only free-streaming takes place until gravitational effects come into play.  Section \ref{sec:gravity} describes in more detail how to further evolve the system at late times, including the gravitational collapse of the minicluster seeds into compact objects. We remark, however, that a full treatment of late-time evolution and gravitational effects is left for a future work.

\section{Axitons}\label{sec:axitons}

\begin{figure}[t]
    \centering
    \includegraphics[width=\columnwidth]{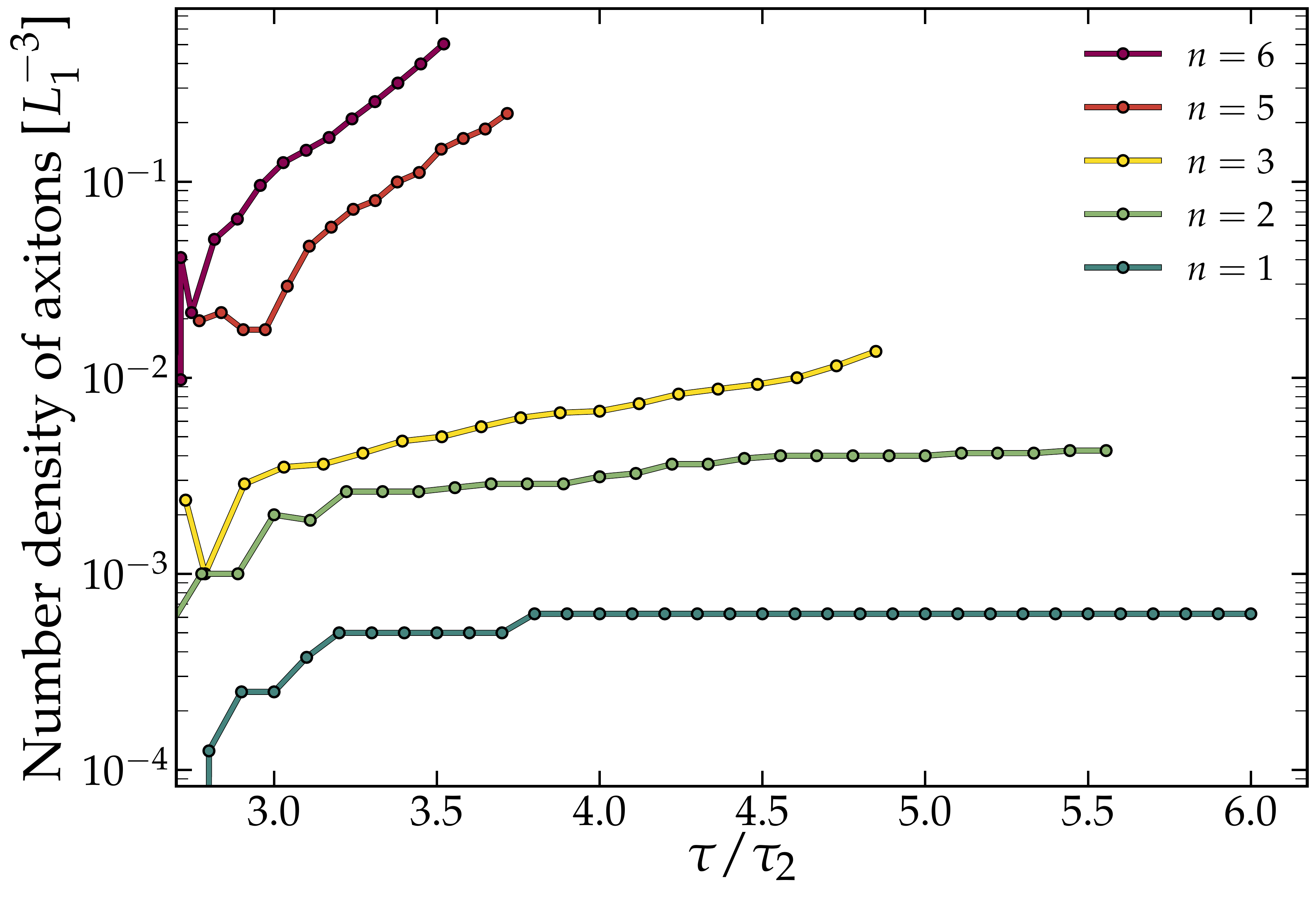}
    \caption{Number density of observed axitons as a function of time in one simulation per value of $\n$. To put each value of $\n$ on a roughly equal footing, we again rescale the time by $\tau_2(\n)$, i.e., the time when the energy densities in the strings and walls coincide [see Eq.~\eqref{eq:tau2}]. Recall that we expect the walls to have collapsed by $\tau/\tau_2 \simeq  2$.}
    \label{fig:Naxitons}
\end{figure}

\begin{figure*}[t]
    \centering
    \includegraphics[height=0.38\textwidth]{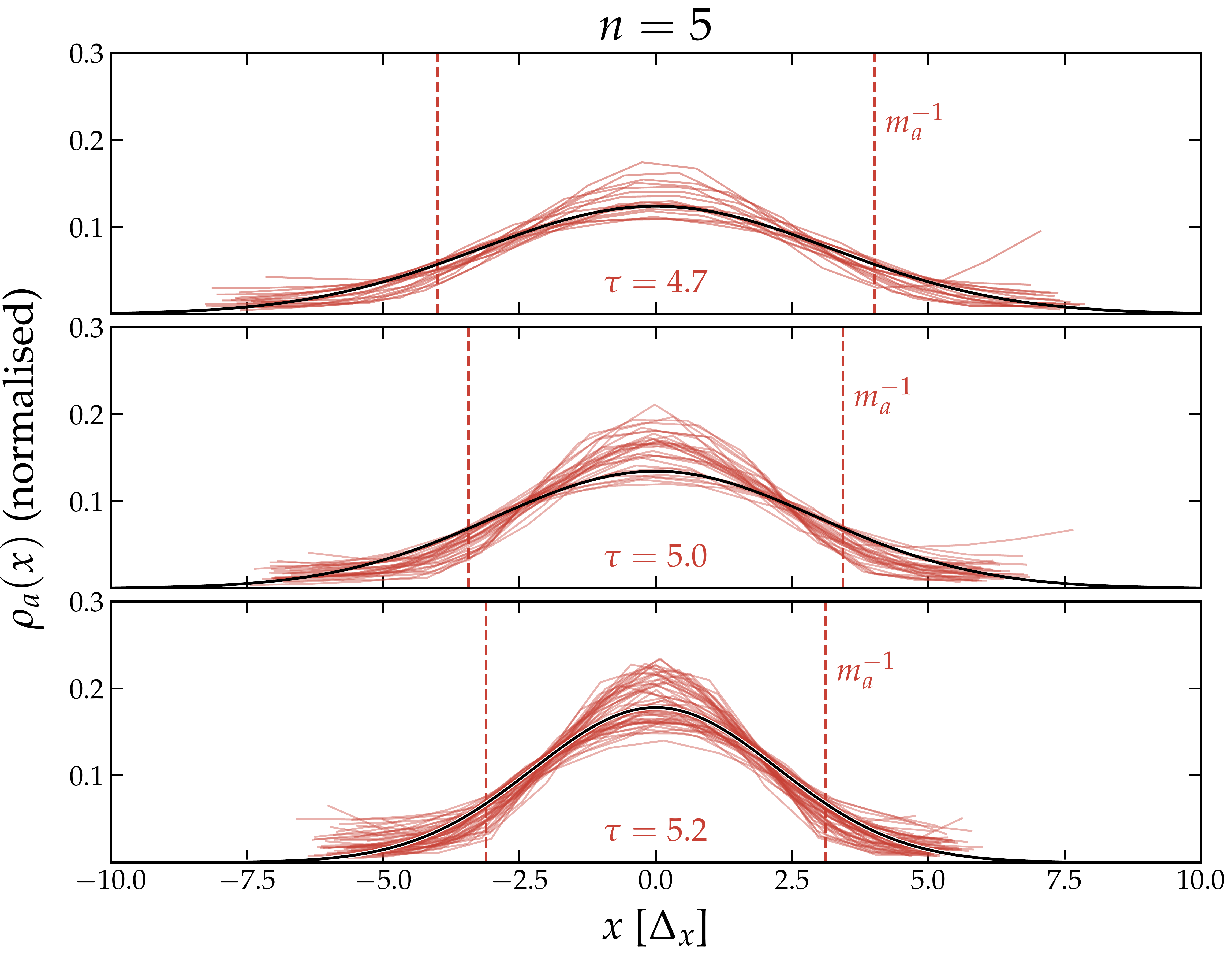}\qquad
    \includegraphics[height=0.38\textwidth]{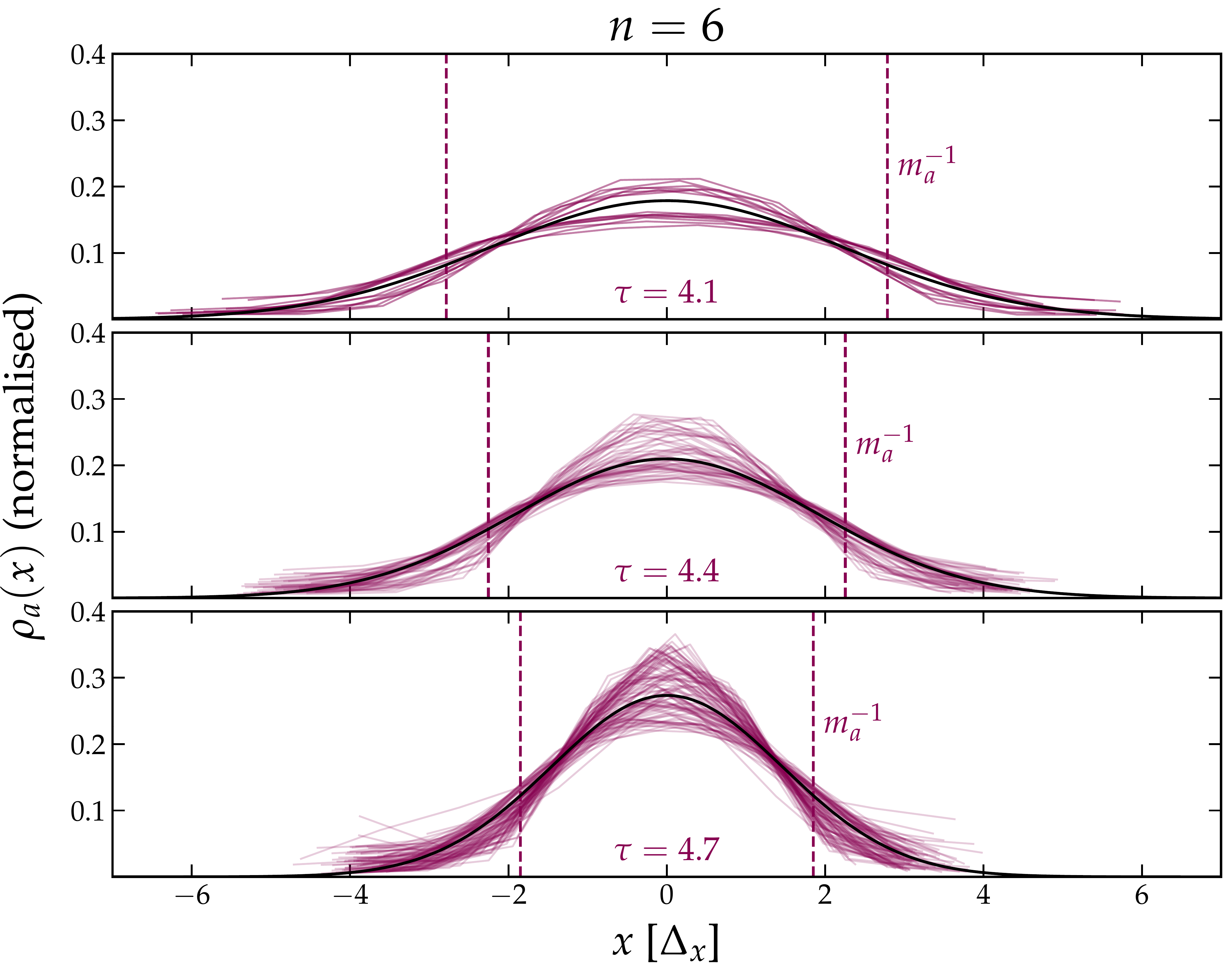}
    \caption{Profiles of the $\sim$100 axitons observed in an $\n=5$ (left panels) and $\n=6$ (right panels) simulation, at three different simulation times (increasing from top to bottom). We indicate the size of the axion Compton wavelength $m^{-1}_a$ at each time, highlighting the fact that the axitons have widths comparable to this scale. Observe particularly in the two latest times (lowest panels) that there are sizeable fluctuations at the edges of some of the profiles.  These are caused by neighbouring axitons, whose typical separations are a fraction of the axion's Compton wavelength. As discussed in Sec.~\ref{sec:axitons}, this happens when the axitons are rapidly multiplying near one another at late times in the $\n=5,6$ cases.}
    \label{fig:axitonprofile_n56}
\end{figure*}

As discussed in Sec.~\ref{sec:dynamics}, towards the end of the domain wall collapse and the subsequent switch to our axion-only simulations, the field develops very small-scale and highly overdense ($\delta_a\gtrsim 100$) configurations called axitons. The conditions leading to the formation of these objects have been described in Ref.~\cite{Vaquero:2018tib}. A Gaussian lump of non-relativistic axions has a critical amplitude~$a^*_{0}$ above which self-interaction dominates the gradient pressure.  When self-interactions dominates, the pressure is not enough to withstand the pull from self-interactions. The lump therefore collapses until potential saturation, happening at $\theta\sim\pi$, when the region has a size similar to $m_a^{-1}$. This is followed by a violent burst of relativistic axions (see Fig.~\ref{fig:axitonprofile_n12}). Interestingly, a similar collapse behaviour has also been observed in the formation of a different but related class of object, the \emph{axion stars}~\cite{Levkov:2016rkk}. Axitons might therefore be thought of as transition regime between the so-called stable ``dilute'' branch of axion stars (which are supported by gravity) and the unstable ``dense'' branch (see, e.g., Fig.~1 of Ref.~\cite{Visinelli:2017ooc} as well as Ref.~\cite{Schiappacasse:2017ham}). In particular, models where $\n>2$ are most likely to develop the aforementioned instability~\cite{Vaquero:2018tib}.

It is not our goal to conduct an exhaustive study of axitons.  But since we have a decent sample of them in some of our simulations, we can draw some rough quantitative conclusions about their abundance and properties for different values of $\n$. We do so by post-processing the simulated axion energy density contrast projected onto a two-dimensional plane,
\begin{equation}
    \delta_{xy} \equiv \frac{1}{\langle \rho_a \rangle} \int \textrm{d}z \rho_a.
\end{equation}
While this means we are missing one dimension to the axitons, we find that they are nevertheless extremely spherical in shape and sparsely distributed  enough to be identified even in a projection.
Our search algorithm looks for closed constant density contours in the $\delta_{xy}$ field with the expected typical size $m^{-1}_a$, that contain densities exceeding some threshold $\delta_{xy}>\delta_{\rm th}$. As an additional cross-check, we also demand that the density contours surrounding the axitons are smooth and circular, i.e., variations in the radial extent of the contour are no larger than the mean radius. We find that thresholds of $\log_{10}(\delta_{\rm th}) =1.5$--2 are optimal for $\n \in [1,6]$.

Figure~\ref{fig:Naxitons} shows the number of axitons found in our simulations as a function of the normalised simulation  time $\tau/\tau_2$ for different cases of $\n$. When $\n$ is large, e.g., $\n=5, 6$, several regions of the field satisfy the instability condition quite early on.   Many axitons quickly form, sometimes even before the final domain walls have collapsed. In these cases, the regions close to the first-formed axitons may subsequently satisfy the instability condition as well, leading to the rapid multiplication of many axitons towards the end of the simulation. Mostly this multiplication entails axitons seeding new axitons nearby. But occasionally a single axiton can appear to divide itself, forming two or more axitons in the process. As a result, the number of axitons in the  $\n=5,6$ cases grows rapidly ($\propto \tau^8,\,\tau^9$ respectively), reaching in excess of 100--200 by $\tau \simeq 4\tau_2$. 
These axitons are also small (relative to $L_1$) and have high density contrasts. We remark, however, that the axiton radii approach $\Delta_x$ towards the final simulation time, so discretisation effects might come into play for the the large $\n$ cases when close to $\tau_f$.

Far fewer regions of the field satisfy the instability condition in the slow mass growth cases, e.g., $\n=1,2$, and the observed numbers of axitons is and \emph{remains} low.
 For $\n = 1$, for example, typically only a handful of axitons appear, and this number plateaus and does not increase further at late times. Lastly, in the $\n=0$ simulations we almost never observe any axiton at all.

Having extracted the axitons  from the field, we can now analyse their profiles. Qualitative
treatments of axitons have previously been presented the literature, using an
exponential~\cite{Levkov:2016rkk,Schiappacasse:2017ham,Visinelli:2017ooc} or
Gaussian~\cite{Vaquero:2018tib} \emph{ansatz} for their profiles under the assumption of spherical
symmetry. We have checked several profiles on our simulations and find that the Gaussian,
exponential, and logistic ($\sech^2$) profiles are all reasonable fits. An attempt to find an
optimal profile is however inconclusive for two reasons. On the one hand, at high values of $\n$ where we
have good statistics (i.e., many axitons), the sizes of the axitons are also very close to the
discretisation scale: only a few grid points cover each structure, making the subtle differences in
the fits to different profiles less meaningful. On the other hand,  the small-$\n$ axitons
are larger and thus can be fitted more accurately to a profile. But, at the same time, they are far less plentiful. As
shown in the upper panels of Fig.~\ref{fig:axitonprofile_n12}, the profile of an individual axiton in the $\n=1$ case actually varies
significantly with time because of the fact that they are quasi-stable and radiate axions. In
particular, the central regions wobble between having flat peaks and sharp ones.  

\begin{figure}[t]
    \centering
    \includegraphics[width=\columnwidth]{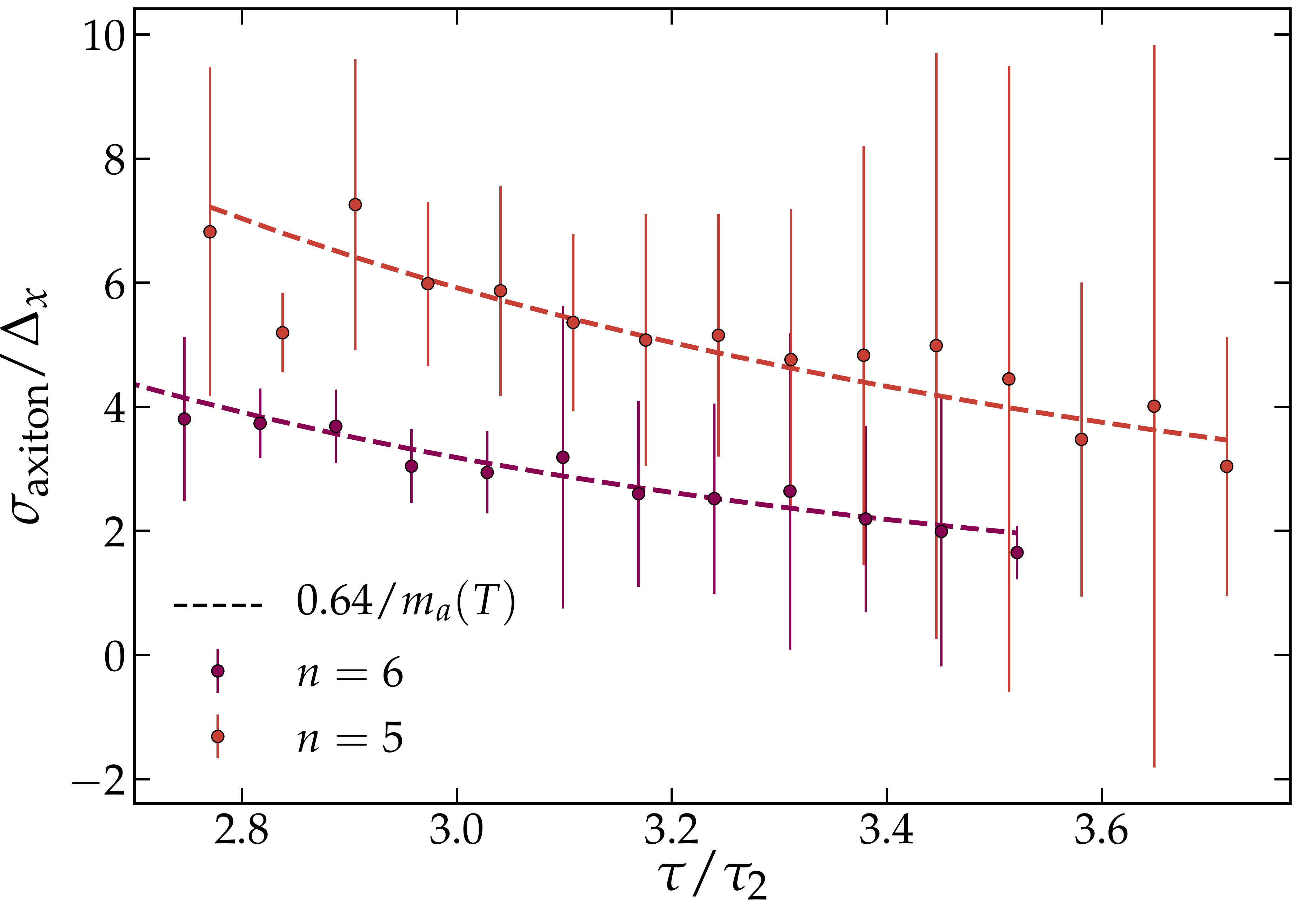}
    \caption{Axiton widths $\sigma_{\rm axiton}$ in units of the discretisation scale $\Delta_x$ as a function of the normalised simulation time $\tau/\tau_2$ for $\n=5,6$.  These widths are determined from 
    fitting a two-dimensional Gaussian profile to the 2D projected density contrast $\delta_{xy}$. Observe that the width decays with the inverse axion mass as expected.  More precisely, we find $\sigma_{\rm axiton}/\Delta_x \sim 0.64 \, m^{-1}_a(T)$ to fit the simulation output quite well.  Notice that at late times in the simulation the error bars become substantially larger. This is because many of the axitons that appear at these late times emerge extremely close to one another---sometimes only a few grid points away---which distorts the fit for a subset of the population.}
    \label{fig:AxitonSizes}
\end{figure}

Focusing on the $\n=5,6$ cases,  we show in Fig.~\ref{fig:axitonprofile_n56}  the 1D profiles of all of our detected axitons at three different times. As expected, the axitons all have comparable shapes with radial widths that are generally around half of the axion's Compton wavelength.  We further quantity the sizes of the axitons as functions of time by fitting a two-dimensional Gaussian to the $\delta_{xy}$ profile with a width in each direction, $\sigma_{x,y}$. We then define $\sigma_{\rm axiton}=\sqrt{\sigma^2_x + \sigma^2_y}$, which we plot as a function of the normalised simulation time $\tau/\tau_2$ in Fig.~\ref{fig:AxitonSizes}.
We observe that this measure of the axiton width is quite well fit by the value $0.64/m_a$ in units of $\Delta_x$ over the range of times shown.

\section{The axion energy spectrum}\label{sec:specshape}

In Sec.~\ref{sec:dynamics} we described qualitatively the dynamics and evolution of the axion field chronologically through our simulations. We will now move to a more quantitative discussion of the resulting distribution of axions. Following the extensive discussions in Ref.~\cite{Vaquero:2018tib} for the case of $\n=7$, we report in this section the features of the axion's \emph{dimensionless power spectrum}, $\Delta^2_a$. This quantity essentially measures the variance of the axion energy density contrast per logarithmic interval in~$k$, i.e., 
\begin{equation}
    \langle\delta_a^2(\mathbf{x})\rangle =\int\mathrm{d\ln} \,k~\Delta^2_a \, ,
\end{equation} 
and can be computed from the Fourier transform of the density contrast%
\footnote{We remark here that in the code we take the Fourier transform of the energy density Eq.~\eqref{eq:endensity} and not the density contrast $\delta_a$.}
\begin{equation}
    \widetilde{\delta}_a(\mathbf{k})=\int \textrm{d}^{3} \mathbf{x} \, e^{i \mathbf{k} \cdot \mathbf{x}} \delta_a(\mathbf{x}) \, 
\end{equation}
via an ensemble average
\begin{mdframed}[linewidth=1.2pt, roundcorner=5pt]
\vspace{-5pt}
\begin{equation}
\Delta^2_a(k)=\frac{k^3}{2\pi^2}\langle\vert\widetilde{\delta}(k)\vert^2\rangle \,. \quad
\label{eq:dimensionlessvariance}
\end{equation}\vspace{-0.15cm}
\end{mdframed} 
 In the following, we use $\Delta^2_a$ to quantify the axion distribution, and pick out some key contributors to the axion abundance during different stages of the evolution.

\subsection{Spectra at early times}

\begin{figure}
    \centering
    \includegraphics[width=0.99\columnwidth]{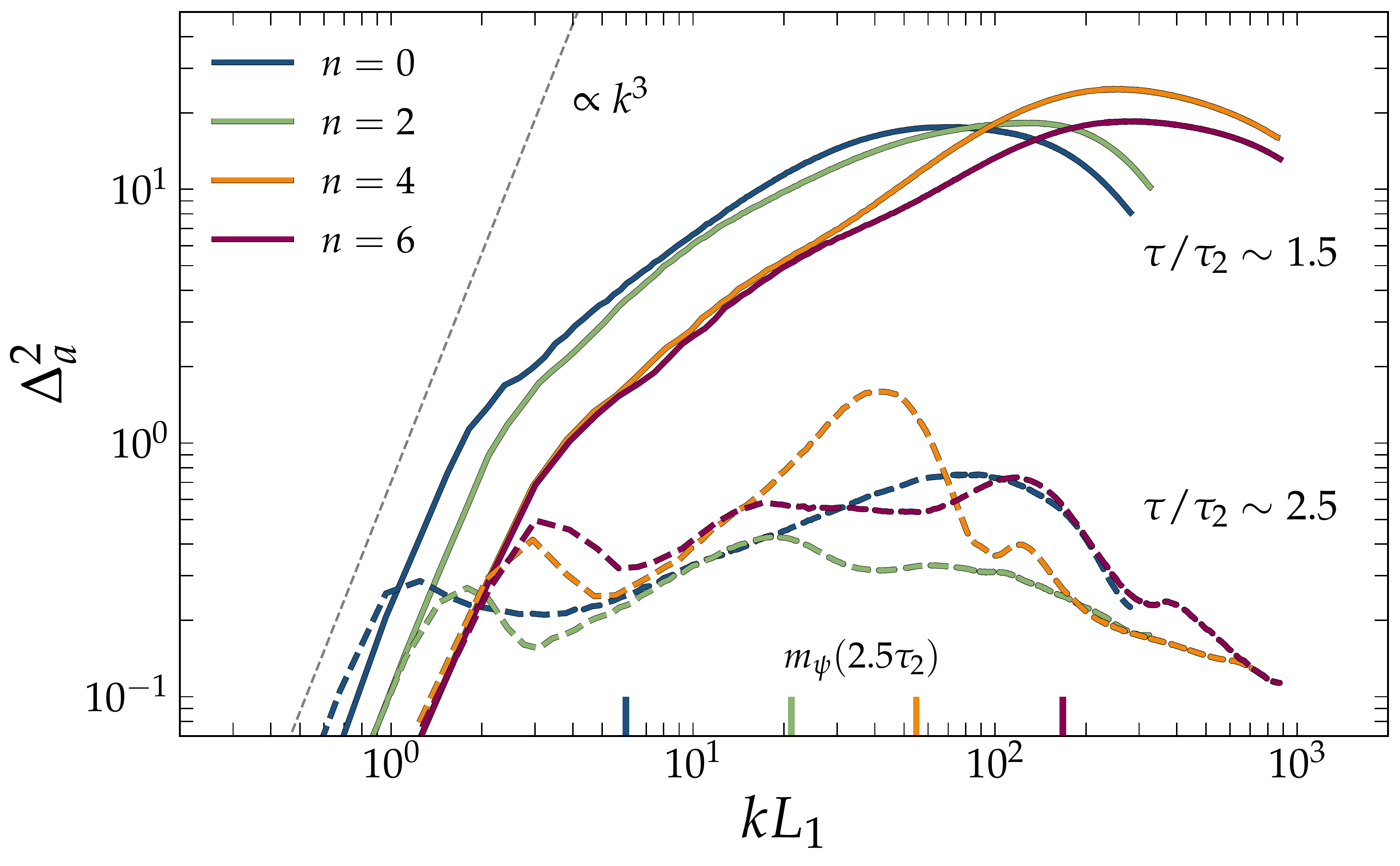}
    \caption{Dimensionless axion power spectrum $\Delta^2_a$ [see Eq.~(\ref{eq:dimensionlessvariance})] as a function of the comoving momentum in ADM units for $\n=0,2,4,6$ at two different times.  In each case, the spectrum shown has been constructed from averaging over ten runs.  We display each spectrum at the rescaled times $\tau/\tau_2\sim 1.5,2.5$, corresponding approximately to the beginning of the network collapse and the post-collapse dynamics, respectively (see Fig.~\ref{fig:scaling}). Observe how the dynamics in the different \n scenarios are comparable when discussed in terms of $\tau_2$. For $\n=2,4,6$ the central peak at the later time is related to the conformal axion mass, $k=m_{\psi}$, while for $\n=0$ the large-$k$ peak is still dominated by relativistic axions emitted by strings.}
    \label{fig:psgeneral}
\end{figure}

In the early stages, i.e., $\tau \lesssim \tau_2$, we expect the power spectrum to be completely dominated by \emph{scaling axions}---axions produced by string decay during the scaling regime. Such axions, being relativistic, will have a momentum distribution between some UV cutoff corresponding to the string width $\sim m_s \sim \Delta_x^{-1}$, and an IR cutoff related to the inter-string separation $k_{\rm IR}\simeq \tau H/\sqrt{\xi}$.%
\footnote{Note that a large $\xi$ can lead to inter-string separations considerably smaller than the Hubble scale.} The expectation is that the energy spectrum of scaling axions is UV-dominated, as most of the axions take a large amount of energy from the strings~\cite{Fleury:2015aca,Gorghetto:2018myk,Vaquero:2018tib}. This can be seen in Fig.~\ref{fig:psgeneral}, where we show the power spectra for several $\n$ cases at a time $\tau<2\tau_2\sim 1.5$, i.e., moments before the string-wall network begins to collapse. The spectra all peak at $k\sim m_s/2 \sim \Delta_x/2$, and drop off again at larger $k$ values due to the UV cutoff. Note that in the smaller-\n cases, these features tend to occur at smaller $k$ values.  This is an artefact of our choice of a larger box size $L_c$ and hence  discretisation scale $\Delta_x$ for these simulations (see Table~\ref{tab:params}).

At the lowest momenta where  large patches of the axion field are essentially uncorrelated, the dimensionless power spectra follow the expectation $\Delta^2_a \propto (k L_1)^3$,
 up to $kL_1\sim 2$--3. These modes represent mostly misalignment-produced axions, and their power decreases over time at a rate that depends on \n, because of free-streaming. Axions that have significant free-streaming velocities act to suppress the uncorrelated part of the power spectrum. The larger the $k$ mode relative to the axion mass under consideration, the more strongly free-streaming acts to remove power.  In the deeply non-relativistic region, i.e., for small velocities $k/m_\psi\ll 1$, the power spectrum freezes, as is evident by the minimal variations in $\Delta_a^2$ between $\tau/\tau_2 \sim 1.5$ and $\tau/\tau_2 \sim 2.5$ at the lowest values of $k$.
 
We generally find that the timescale $\tau\simeq 1.5\tau_2$ provides an upper bound on the power $\Delta^2_a$ at the IR cutoff in all cases except $\n=0$. For $n\geq 6$, the IR cutoff does not change considerably with time, as can be seen in Fig.~\ref{fig:psgeneral}, where $\Delta^2_a(k_{\rm IR})\sim 10^{-1}$ throughout most of the evolution. This is not surprising, as such low momentum modes are frozen right after the collapse. For $\n=0$ on the other hand, the largest value of $\Delta^2_a(k_{\rm IR})$ is achieved at a later stage in the evolution, at $\tau/\tau_2\sim 2$, due to the increased amount of the free-streaming of relativistic axions.

The story at high momenta is very different. After the collapse of the network begins $\tau/\tau_2 \gtrsim 1$, the spectrum loses power as the mildly relativistic axions emitted can free-stream and wash-out perturbations at the very small scales. This can been seen in Fig.~\ref{fig:psgeneral}, where the broad high-$k$ peak in the spectrum has disappeared completely by $\tau\sim 2.5\tau_2$. This is the case for all values of $\n$ with the exception of $\n = 0$, which maintains a broad high-$k$ peak even at $\tau/\tau_2>2.5$.

We observe also in Fig.~\ref{fig:psgeneral} that in the $\n>0$ cases, the spectrum changes shape and develops a new peak on scales $k\sim m_a^{-1}$, a phenomenon that can quite clearly be attributed to the decay of the domain walls and the subsequent appearance of axitons: since the axion Compton wavelength $\lambda_c\sim \pi/m_a$ sets the domain wall width~\cite{PhysRevLett.48.1156} [see Eq.~\eqref{eq:walldensity}], we generally should expect to see non-linear features in the power spectrum around the corresponding momentum. This remnant peak from the wall collapse is shifted towards higher $k$ as we increase $\n$, since for larger values of $\n$ the axion mass grows faster and is therefore larger at the times shown. As we move forward in time all of these peaks will move gradually to the right as the axion mass grows, and for the large $\n$ cases in particular, the peak will grow substantially in amplitude as more and more axitons are produced. See, e.g., Fig.~9 of Ref.~\cite{Vaquero:2018tib}. 

\subsection{Spectra at late times}

\begin{figure}
    \centering
    \includegraphics[width=\columnwidth]{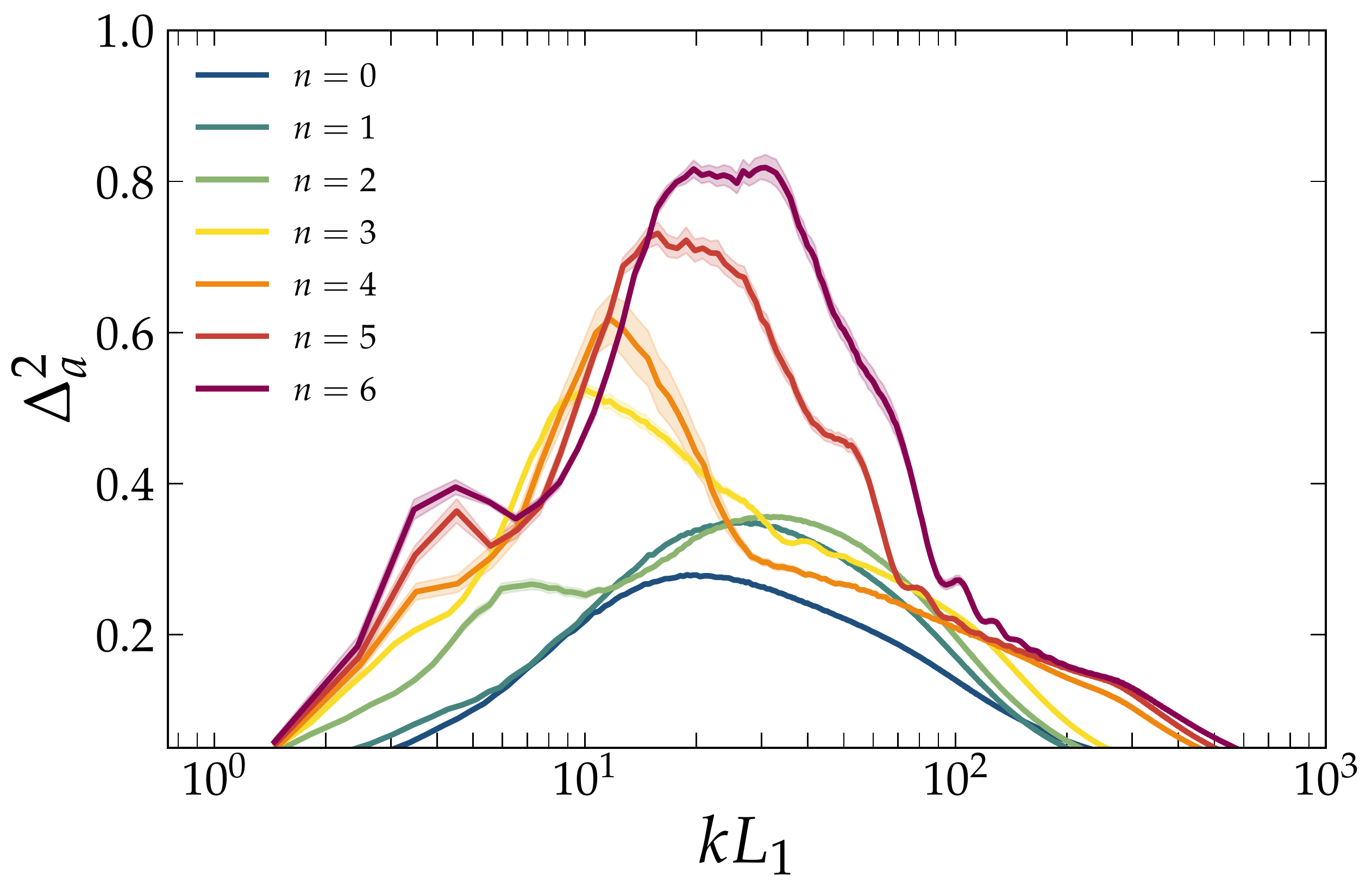}
    \caption{Dimensionless power spectrum of the axion energy density contrast $\Delta^2_a$, defined in Eq.~(\ref{eq:dimensionlessvariance}), as a function of the axion's momentum in units of $1/L_1$. We show seven power spectra, one for each value of $\n$ considered in this work, ranging from the generic ALP case (lowest, blue line), to a case close to the canonical QCD axion (highest, red line). The band around each spectrum encodes the standard error collected from ten simulations per value of $\n$.}
    \label{fig:pswkb}
\end{figure}

Figure~\ref{fig:pswkb} shows the power spectra $\Delta^2_a(k)$ for all \n scenarios at $\tau_{\rm ad}(\n)$, after applying the WKB approximation as described in Sec.~\ref{sec:adiabatic}. Qualitatively, between $\tau_f$ and $\tau_{\rm ad}$ the high-$k$ part of the spectrum gets significantly reduced because of the absence of self-interactions and the disappearance of the axitons.  Quantitatively, however, we caution that the adiabatic approximation may not be reliable at the highest-$k$ end of the spectrum, i.e., $kL_1\gtrsim 30$--$100$ depending on the value of \n, where the modes have large velocities $k/m_\psi \gtrsim 0.1$.%
\footnote{We stress that in the $\n=0$ scenario, only modes up to $kL_1\sim 50$ have transitioned to non-relativistic regime, and that only modes $kL_1\lesssim 10$ have small velocities $k/m_{\psi}$.} 
It is clear that even a small change to the axion mass growth rate index leads to substantial and non-trivial changes to the spectrum. Our largest \n simulation shows a peculiar shape with a maximum value at $kL_1\sim 20$--40, quite close to the $\n=7$ result of Ref.~\cite{Vaquero:2018tib}, plus a smaller, second peak at larger scales $kL_1\sim 4$. The main peak height decreases with decreasing $\n$, and its location also appears to shift to larger scales; for $\n=3,4$ we find the main peak at $kL_1\sim 10$. On the other hand, the $\n\leq 2$ scenarios produce power spectra that are much smoother in shape, with a broad peaks and generally much less power at large scales.
Note that all spectra maintain a $k^3$ slope in the low-$k$ region, and we find $\Delta^2=C(kL_1)^3$ with  $C\simeq 0.03$--0.05 to provide a good  description.

\section{Dark matter}\label{sec:darkmatter}
Having applied the adiabatic approximation once the bulk of the axions are deep in the non-relativistic regime, we can now proceed to calculate their abundance. To do this we assume no additional entropy injection at later times, so that the number of axions per comoving volume is conserved, $n_a/s={\rm const.}$, where $s$ is the entropy density. Assuming entropy conservation, this fact allows us to estimate the total relic abundance of DM, and therefore predict windows of ALP parameter space that are consistent with observed cosmology. 

Our goal here is to compare the efficiency of axion production relative to the yield from the standard analytic calculation for the zero-momentum mode misalignment, with a particular focus on the $\n$-dependence. Additionally, given that we have performed simulations for $\n$ up to 6, we may be able to extrapolate our results to estimate the QCD axion abundance without the costly grid sizes needed to simulate large $\n$ physically.

A similar estimate of the axion production efficiency was carried out by Ref.~\cite{Klaer:2017ond} for the QCD axion ($\n\in[6,8]$), and for $\n=2$. They found that when $\n=2$, axions are produced in greater abundance thanks to the slow decay of the topological defects. Recently, Ref.~\cite{Chaumet:2021gaz} confirmed this observation by comparing the $\n=0$ and $\n=7$ scenarios in an $O(N_d)$ theory with $N_d\geq 3$. Despite the fact that the physical case should correspond to $N_d=2$, this result confirms the intuition that models with slow mass growth are generally more efficient at producing axions. In the following, we present our estimation of the axion production efficiency for the $U(1)$ model, including for the first time the case $\n=0$.

\subsection{Number density}

Naively, we expect the number of axions produced per unit comoving volume via misalignment to be $n_a \sim H_1 f^2_a R_1^3$, although the precise numbers cannot be estimated easily analytically. In our simulations, however, we can just count the number of axions per comoving volume, and this number will include contributions from topological defects as well as non-linearities~\cite{Fleury:2015aca}. To do this, we compute~\cite{Vaquero:2018tib}
\begin{equation}\label{eq:na}
    \frac{n_a}{m_1f^2_aR^3_1}=\frac{1}{c_1(\n)}\int \frac{k^2{\rm d}k}{2\pi^2} \mathcal{N}(k) \, ,
\end{equation} 
where the integrand is the angle-averaged \emph{occupation number}, an adiabatic invariant that can be written as
\begin{equation}
\mathcal{N}(k)=\left\langle\frac{1}{2w_k}\vert\partial_{\tau}\tilde{\psi}\vert^2+\frac{w_k}{2}\vert\tilde{\psi}\vert^2\right\rangle_{\vert \mathbf{k}\vert=k}.
\end{equation} 
Note that we have also used the relation~\eqref{eq:c1}
 to ensure that we keep track of the different $t_1$ criteria for different $\n$ scenarios.

\begin{figure}
    \centering
    \includegraphics[width=0.98\columnwidth]{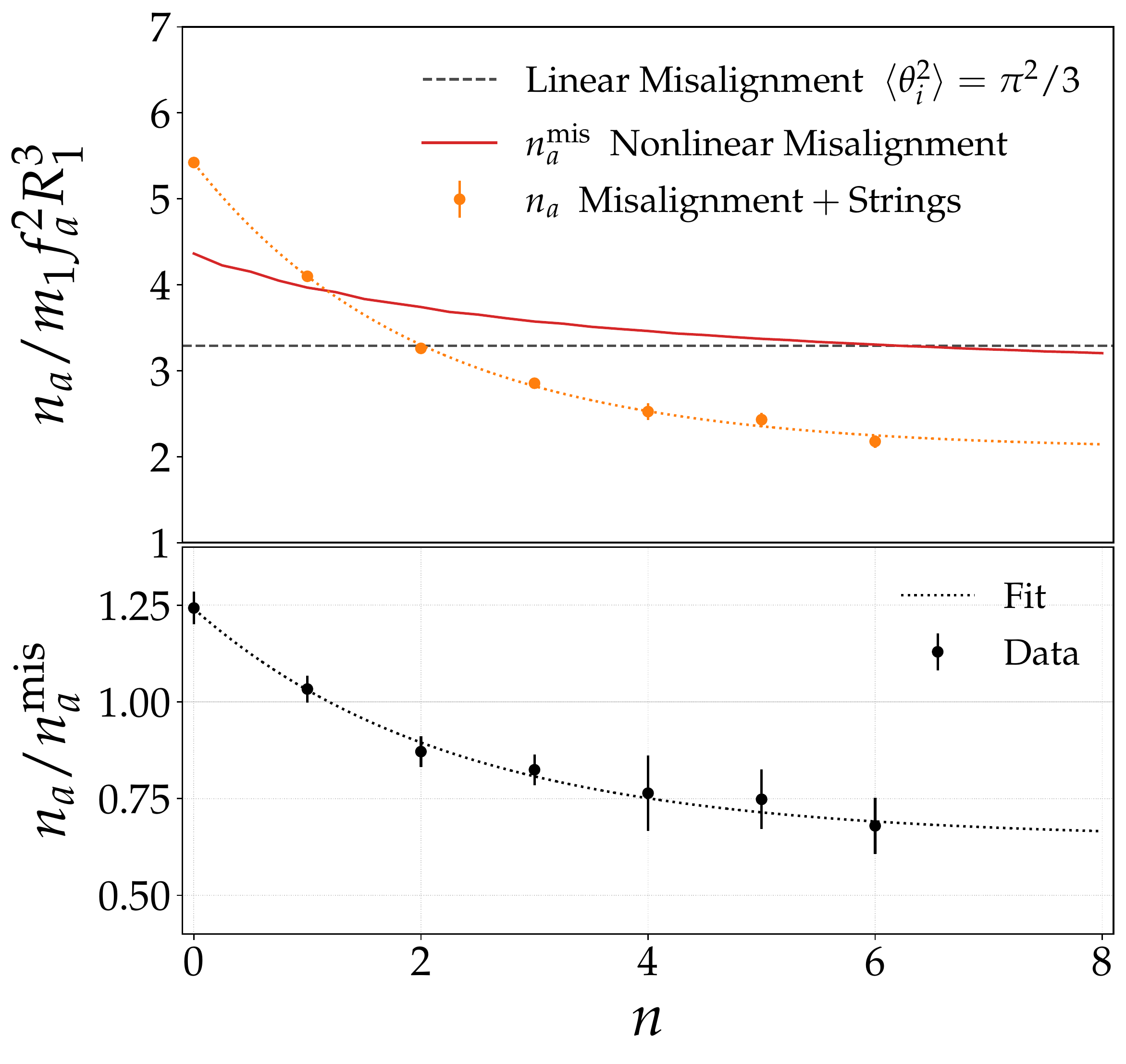}
    \caption{Comoving axion number density in the adiabatic regime as a function of $\n$. In the upper panel, the orange points with error bars represent our simulation results, which are well fitted by Eq.~\eqref{eq:nafit}, indicated by the orange line. 
    The red line denotes the average number density from nonlinear misalignment, constructed from the solution of Eq.\eqref{eq:misnum}. 
    For additional clarity we compare these results with the statistical prediction from the \emph{linear} misalignment calculation (gray dashed line), that is based on an averaged initial angle of $\langle \theta^2_i\rangle=\pi^2/3$. The bottom panel displays the production efficiency, i.e., the number density obtained from our simulations relative to that from the nonlinear misalignment calculation, $n_a/n_a^{\rm mis}$. 
    }
    \label{fig:dm}
\end{figure}

When we track the axion number density over time we find that it reaches a peak value at an equivalent point in the evolution for each $\n$ simulation at around $\tau/\tau_2\sim 1.25$, followed by a steep decrease to a constant value. This plateauing of the number density occurs after around $N_{\rm osc}\sim 10$ background (zero-mode) field oscillations around the potential minimum, with lower values of \n plateauing at higher number densities. Our final number density as a function of \n is shown in the upper panel of Fig.~\ref{fig:dm}. We see that it is well fit by the function
\begin{equation}
    \frac{n_a}{m_1f^2_aR_1^3}=a(1+be^{-c~\n})+d, \label{eq:nafit}
\end{equation} 
with $a=4.03, b=0.83, c=0.50$, and $d=-1.96$.

\subsection{Nonlinear misalignment and efficiency}\label{sec:nonlinmis}

Now that we know the total number density of axions at the end of our simulations, we can compare this with the number density expected from the standard zero-mode misalignment calculation. We treat the full nonlinear misalignment in a way consistent with our $t_1$ criterion, by numerically solving the $\n$-dependent equation
\begin{equation}
    \partial_{\hat{t}}^2{\theta}+\frac{3}{2\hat{t}}\partial_{\hat{t}}{\theta}+\left(\frac{c_1(\n)}{2}\right)^2\hat{t}^{\n/2}\sin(\theta)=0\,, \label{eq:misnum} 
\end{equation}
which has the Laplacian term set to zero.  Observe that the time derivatives are with respect to $\hat{t}=t/t_1$, and consequently the $c_1(\n)$ coefficient enters the equation with a different prefactor compared to Eq.~\eqref{eq:c12linear}.  We solve Eq.~\eqref{eq:misnum} from $\hat{t}_i$ to $\hat{t}_f$, given respectively by
\begin{equation}
    \hat{t}_i=10^{-3},~~~~~~~\hat{t}_f=(N_{\rm osc}\pi(\n+4)/2)^{4/(\n+4)}\, , \label{eq:misnum1}
\end{equation} 
where the final time $t_f$ corresponds to $N_{\rm osc}$ field oscillations around the potential.
We choose $N_{\rm osc}=8$ and solve for a range of initial angles $\theta_i \equiv \theta(\hat{t}_{i})\in [0,\pi)$. The misalignment number density can then be calculated from this solution via
\begin{equation}
    \frac{n_a^{\rm mis}}{m_1f^2_a R_1^3}=\frac{4}{\pi c_1(\n)^2}\int_0^{\pi}{\rm d}\theta\frac{\theta^2(\hat{t}_{f})}{\theta^2_i}\frac{\rho_{\theta}(\hat{t}_{f})}{m(\hat{t}_{f})}R(\hat{t}_{f})^{3}, 
\end{equation}
where
\begin{equation}
    \rho_{\theta}(\hat{t})=\frac{1}{2}\left[\left(\frac{{\rm d}\theta}{{\rm d}\hat{t}}\right)^2+\left(\frac{c_1(\n)\,\hat{t}^{\n/4}}{2}\right)^2(1-\cos\theta)\right]\,
\end{equation} 
represents the energy density stored in the $\theta$ field at time~$\hat{t}$.
 
 The value of $n_a^{\rm mis}$, in units of $m_1f^2_aR_1^3$, as a function of $\n$ is shown alongside our simulation results in Fig.~\ref{fig:dm}. We also calculate the efficiency by taking the ratio of these two results, $n_a/n_a^{\rm mis}$,  shown in the lower panel of Fig.~\ref{fig:dm}. We see that the most efficient case is $\n=0$, where the decay of the strings and walls produces an additional 25\% more axions over the misalignment-only scenario. On the other hand, the case of $\n=6$ reveals a low efficiency at only $n_a/n_a^{\rm mis}\simeq 0.68\pm 0.07$. 

The general trend that lower values of $\n$ results in more axions is in broad agreement with the findings of Refs.~\cite{Klaer:2017ond,Chaumet:2021gaz}. However, quantitatively, our results at small string tensions $\kappa\lesssim 8$, show that only in the $\n=0,1$ cases is it possible to obtain yields higher than the misalignment estimation. On the other hand, Ref. \cite{Klaer:2017ond} performed simulations adding string tension and computing the abundance for QCD axion physical values $\kappa\sim 60-70$. While for large $n$ the effect of added string tension is modest, of order 30\% in the DM density, for small $n$ cases it is dramatic: for added-tension
simulations the efficiency is a few times higher than the misalignment estimation.\footnote{G. Moore, private communication.}

\subsection{Relic abundance and outlook}

\begin{figure*}
    \centering
     \includegraphics[width=0.49\textwidth]{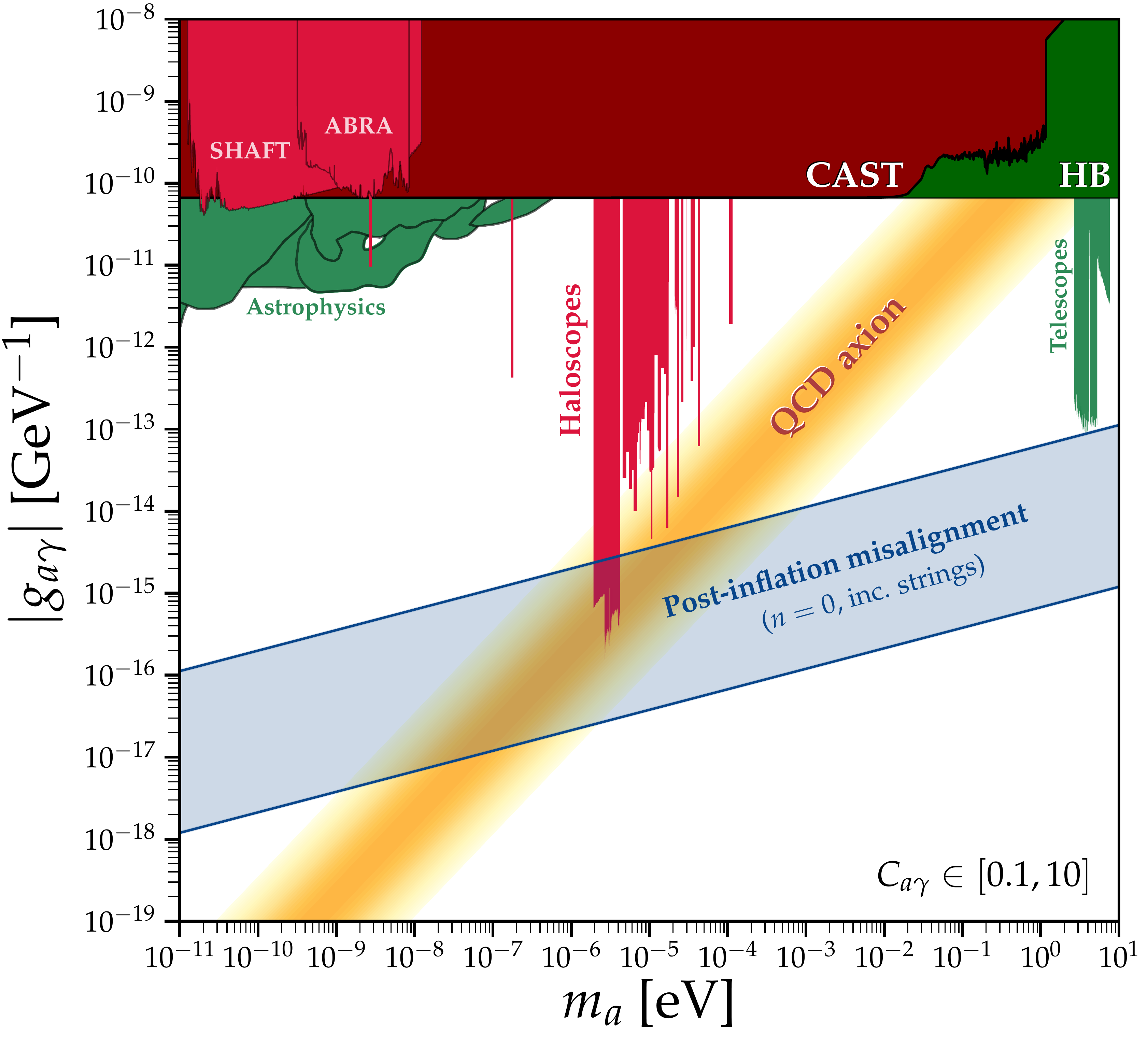}
    \includegraphics[width=0.49\textwidth]{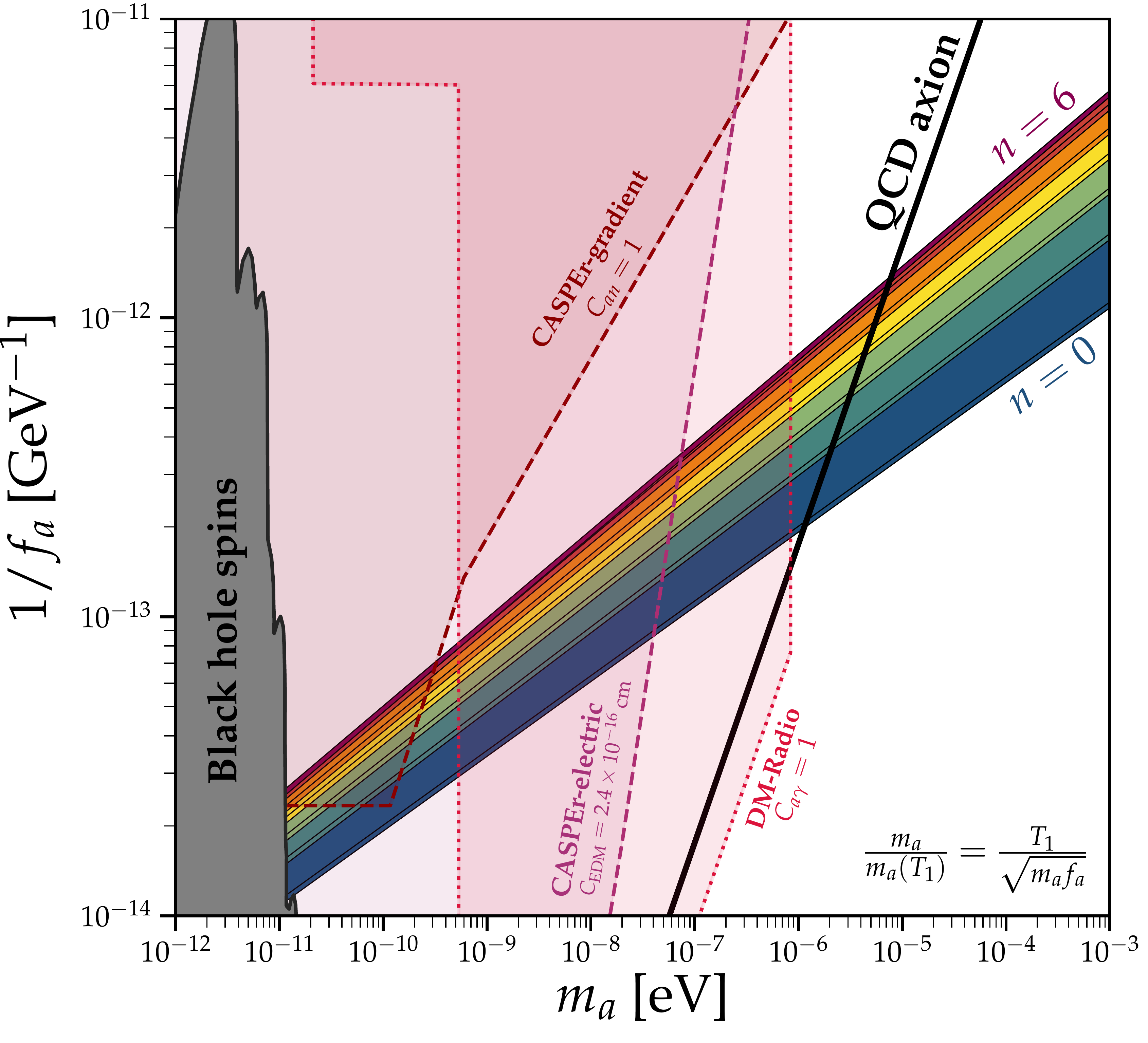}
    \caption{Range of temperature-dependent ALP masses and couplings that can comprise the totality of the observed cosmological DM abundance. In the left panel we display the familiar photon-coupling parameter space, where the blue band indicates a possible window of DM ALP couplings for a range of arbitrary dimensionless photon couplings, $C_{a\gamma} \in [0.1,10]$, in the $\n=0$ case. We also show existing constraints from DM haloscopes~\cite{DePanfilis,Hagmann,Crisosto:2019fcj,ADMX:2018ogs,Bartram:2021ysp,Asztalos2010,ADMX:2018gho,ADMX:2019uok,ADMX:2021abc,Ouellet:2018beu,Salemi:2021gck,Lee:2020cfj,Jeong:2020cwz,CAPP:2020utb,Devlin:2021fpq,Grenet:2021vbb,HAYSTAC:2018rwy,HAYSTAC:2020kwv,McAllister:2017lkb,Alesini:2019ajt,Alesini:2020vny,CAST:2021add,Gramolin:2020ict}, astrophysical photon-axion oscillation~\cite{HESS:2013udx,Payez:2014xsa,Fermi-LAT:2016nkz,Li:2020pcn,Dessert:2020lil,Reynolds:2019uqt}, telescope searches for DM ALPs in galaxies~\cite{Regis:2020fhw,Grin:2006aw}, the horizontal branch (HB) star cooling~\cite{Ayala:2014pea}, and the bound on solar axions from CAST~\cite{CAST:2007jps,CAST:2017uph}. In the right panel we show the window in the ($m_a$,$f^{-1}_a$)-parameter space
    for all $\n$ values [assuming the mass growth enhancement factor~\eqref{eq:en}], as well as projections for three possible future experiments: CASPEr-electric~\cite{Aybas:2021nvn,JacksonKimball:2017elr} (via the electric dipole moment coupling), CASPEr-wind~\cite{Garcon:2019inh} (via the ALP-nucleon coupling),  and DM-Radio~\cite{DMRADIO} (via the photon coupling). We also show the bounds from black hole superradiance~\cite{Mehta:2021pwf}. Data and references for all of the limits shown here can be found in Ref.~\cite{AxionLimits}.}
    \label{fig:dm1}
\end{figure*}

With an estimate of the number density of axions in hand, we can now fix some physical parameters for the axion and finally estimate the present-day relic abundance. Using entropy conservation and Eq.~\eqref{eq:na}, we find the present-day energy density in axions to be 
\begin{widetext}
\begin{equation}\label{eq:rhotoday}
    \rho_{a,0}(\n)\simeq 0.15~\frac{{\rm eV}}{{\rm cm}^3} ~\theta^2_{\rm eff}(\n)\,{\cal{F}}_{70}(T_1)\left(\frac{c_1(\n)}{1.6}\right)^{3/2}\left(\frac{m_a}{m_1}\right)^{1/2}\left(\frac{m_a}{\upmu{\rm eV}}\right)^{1/2}\left(\frac{f_a}{10^{12}~{\rm GeV}}\right)^2~,
\end{equation} 
\end{widetext} 
with ${\cal{F}}_{70}(T_1)=(g_{*}(T_1)/70)^{3/4}(g_{*s}(T_1)/70)^{-1}$. The result of our simulations is encapsulated in the
\emph{effective} initial angle, $\theta^2_{\rm eff} \equiv n_a/m_1f^2_aR_1^3$, i.e., the angle one would need to choose in order to correct the naive abundance estimate to reproduce the result that includes topological defects. The numerical values we obtain for $\n\in [0,6]$ are
\begin{equation}
    \theta_{\rm eff}
    =\{2.33,2.02,1.81,1.69,1.59,1.56,1.48\} \, ,
\end{equation} taken from an average over 10 simulations per value of \n, and with roughly 5\% uncertainties.

The simplest case to estimate is the $\n=0$ scenario, since the characteristic temperature $T_1$ and hence $m_1 \equiv m_a(T_1)$ depends only on the mass $m_a$, which stays constant in time.   Then, evaluating Eq.~\eqref{eq:rhotoday} for $m_1 = m_a$ we find
\begin{equation}
    \Omega^{\n=0}_ah^2\simeq 0.019~{\cal{F}}_{70}(T_1)~\left(\frac{m_a}{\upmu{\rm eV}}\right)^{1/2}\left(\frac{f_a}{10^{12}~{\rm GeV}}\right)^2 \, 
    \label{eq:omegatodayn0}
\end{equation}
for the axion energy density relative to the critical density, $\Omega_{a,0}~\equiv~\rho_{a,0}/\rho_{\rm crit}$. If we further assume all DM to be made of cold axions and fix $ \Omega^{\n=0}_ah^2$ to the observed DM density, a simple relation between $m_a$ and $f_a$ can be straightforwardly established from Eq.~\eqref{eq:omegatodayn0}. A conservative lower bound on the abundance in this case was calculated in Ref. \cite{Gorghetto:2021fsn}, from the high resolution simulations of \cite{Gorghetto:2020qws} and under the assumption that the axion spectrum becomes IR dominated at $\kappa\gg  1$.

In order to obtain a similar relation for the $\n>0$ scenarios, we need to make an additional assumption about the factor $\sqrt{m_a/m_1}$ in Eq.~\eqref{eq:rhotoday}. In other words, how long should the axion mass continue to grow? For a generic ALP model, the mass growth could be attributed to a hidden SU($N$) group that condenses at some scale $\Lambda$. In this case, the enhancement factor could be simply estimated as~\cite{Arias:2012az,Blinov:2019rhb}
\begin{equation}
    \sqrt{\frac{m_a}{m_1}}\sim \left(\frac{T_1}{\Lambda}\right)^{1/2}, \label{eq:en}
\end{equation} 
with $T_1$ given by Eq.~\eqref{eq:t1alps} and $\Lambda=\sqrt{m_a f_a}$. We will adopt this assumption when showing results for our $\n>0$ scenarios, but emphasise that this is just one possible choice. Alternatively, we could simply choose some arbitrary time to saturate the axion mass growth. We observe in our simulations that the comoving axion number density $n_a$ tends to a constant value towards the end of our simulations. So as far as the DM abundance is concerned we should be free to identify any time after the final simulation time $\tau_f$ to be the saturation time $\tau_*$ [see Eq.~\eqref{eq:axionmass_T}], as long as we can continue to assume $n_a(\tau)=n_a(\tau_f)$ for all $\tau\geq \tau_f$. This option clearly comes with a lot more freedom though, hence why we choose the simpler and more restrictive case described above.

We can now derive windows of viable ALP models by fixing $\Omega_{a,0}h^2$ to the DM density measured by the Planck mission, $\Omega^{\rm Planck}_{\rm DM} h^2 = 0.12$~\cite{Aghanim:2018eyx}, and rearranging Eq.~\eqref{eq:rhotoday} for the axion decay constant $f_a$ under the assumptions discussed above. The right panel of Fig.~\ref{fig:dm1} shows the decay constant $f_a$ versus the axion mass $m_a$ for all  $\n \in [0,6]$ considered in this work, keeping in mind that only the generic temperature-independent ALP ($\n =0$) is free from the assumption~\eqref{eq:en} for the axion mass growth enhancement. We display also in the left-hand panel of 
Fig.~\ref{fig:dm1} the $\n=0$ prediction in terms of the ALP-photon coupling, defined in analogy to the QCD axion as $g_{a\gamma} \equiv \alpha C_{a\gamma}/(2\pi f_a)$.  Note that some degree of model dependence remains, as the dimensionless coupling constant $C_{a\gamma}$ is arbitrary for ALP models; the blue band indicates a range of possible values $C_{a\gamma}\in[0.1,10]$.

The intention behind the left panel of Fig.~\ref{fig:dm1} is to give the expected range of parameters for post-inflationary ALPs in a familiar context, and to show the relationship between our prediction and the broader landscape of searches (see figure caption). The right panel, on the other hand, highlights some experiments that might be able to probe the various models in the future. Taking the (arguably optimistic) projections set by these collaborations at face value, we can expect CASPEr-electric and DM-Radio to cover everything below $\sim \upmu$eV, and CASPEr-gradient to reach only the very lightest masses if the mass-growth enhancement factor $\sqrt{m_a/m_1}$ is large enough.%
\footnote{We define the axion-EDM coupling to be $g_{an\gamma} = e C_{\rm EDM}/f_a$ with $C_{\rm EDM} = 2.4\times 10^{-16}$~cm~\cite{Pospelov:1999mv}, and the axion-nucleon coupling as $g_{an} = C_n m_n/f_a$, with an arbitrary $C_n$.} 
Unfortunately, the high-mass end of our predicted ALP DM band implies significantly weaker couplings than the QCD axion, which would make the search for these models considerably more challenging. It is possible that for large-enough values of $C_{a\gamma}$ some future telescope-based searches similar to  Refs.~\cite{Regis:2020fhw,Grin:2006aw} looking for heavy ALPs decaying inside galactic halos may reach these models, though this is not clear at present. In contrast, the low-mass end will be readily probed as experiments continue to make progress towards their primary goal: the QCD axion.

\section{Miniclusters}\label{sec:gravity}

As we approach matter-radiation equality, the small-scale power in the axion field seen at the end of our simulations will start to form the first gravitationally bound structures of DM called axion miniclusters. We remark that, although the appearance of this structures is seeded by large $\mathcal{O}(1)$ overdensities in the axion energy field, they are not directly related to the presence of axitons (the latter are expected to disappear way before any gravitational collapse). In this final section we would like to predict the properties of the eventual miniclusters, using only the information we possess about the spatial clustering of their early seeds and the power spectrum $\Delta^2_a$ after the WKB approximation, while \emph{without} performing any N-body simulations. 

As pointed out in Ref.~\cite{Vaquero:2018tib}, and confirmed by the spectra shown in Fig.~\ref{fig:pswkb}, the presence of strings and domain walls in the initial conditions leads to substantial power on scales much smaller than $L_1$. This modifies somewhat the conventional wisdom that miniclusters form as a result of spatial fluctuations on the typical scale $L_1$, corresponding to an average mass $M_1 = 4 \pi/3 \langle\rho_a\rangle L^3_1$---about $10^{-12} M_{\odot}$ for typical QCD axion scenarios, but potentially much smaller for general ALPs.  Rather, the power spectrum suggests we should expect a substantial population of collapsed objects with masses much smaller than $M_1$.
To estimate the mass distribution of these objects, or the minicluster \emph{halo mass function} (HMF), we follow the arguments of Ref.~\cite{Kolb:1994fi}, who first studied the formation of miniclusters by numerically computing the non-linear spherical collapse of large overdensities in a radiation background. We also apply a modified version of the Press-Schechter (PS) formalism, wherein the gravitational collapse of some substructure can take place before matter-radiation equality.

In the standard PS prescription, the amount of cold matter in gravitationally bound objects of a particular size $\sigma_s$ at a given time can be read off the cumulative probability distribution of smoothed density fluctuations  $\delta_a(\sigma_s)$ larger than some threshold value $\delta_c$.
Here, the smoothed density contrast $\delta_a(\sigma_s)$ is constructed by filtering a linearly-evolved (under gravity) density field over $\sigma_s$, which we call the \emph{smoothing length}. The collapse threshold $\delta_c$ is usually established from the spherical collapse model, and has  the well-known value $\delta_c = \delta_c^* \simeq 1.686$ for standard cold DM during matter domination. Here, we use~\cite{Kolb:1994fi,Ellis:2020gtq}
\begin{equation}
    \delta_c(R)=\frac{1.686}{3}\frac{2+3R/R_{\rm eq}}{R/R_{\rm eq}},\label{eq:deltac}
\end{equation}
in order to include collapses before matter-radiation equality.

We implement a real-space top-hat filter to smooth our simulated density contrast field $\delta_a(x)$, after applying the adiabatic approximation, for a set of $N_{\sigma}=4\log_2(N)$ different values of $\sigma_s$ spanning the grid-spacing $\Delta_x$ and the box size~$L_c$. From this we build the functional integral for the cumulative mass, defined as
\begin{equation}
    {\cal{M}}(\sigma_s,R)=L_c^3\langle\rho_a\rangle\int_{\delta_c(R)}p(\delta_a,\sigma_s)(1+\delta_a){\rm}\delta_a \, .
\end{equation}
To convert this to an HMF, we note that the smoothing length $\sigma_s$ can be related to the mass of the eventual minicluster via $M = 4 \pi/3 \langle\rho_a\rangle \sigma^3_s =
M_1(\sigma_s/L_1)^3$. This then allows us to define the HMF as \begin{equation}
    \frac{{\rm d}n}{{\rm d}\log M}=\frac{1}{L^3_c}\frac{{\rm d}\cal{M}}{{\rm d}M},\label{eq:hmf}
\end{equation} 
which represents the comoving number density of miniclusters per logarithmic mass interval.

In this section we implement the process described above on a $\n=0$ simulation, as well as a QCD axion simulation with $\n=6.7$. The latter entails performing a new simulation on a larger grid of $N=4096^3$ points in order to avoid excessive tilting of the saxion potential. The other simulation parameters set to $L_c=6L_1$, $\delta_{\rm core}^{-1}=1.25$, and $\tau_f=4.5$ following the same arguments presented in Sec.~\ref{sec:simulation}. Then, we fix $m_a=10^{-4}$ eV, and in the case of $\n=0$, we additionally fix the value of $f_a$ via Eq.~\eqref{eq:rhotoday} by demanding that the axion abundance matches the Planck DM density measurement. These assumptions in turn fix the characteristic temperature~$T_1$ and consequently the characteristic scale $L_1$ to be
\begin{align}
    L_1^{(\n=0)}&\simeq 1.6\times 10^{-4}~{\rm pc}, \\
    L_1^{(\n=6.7)}&\simeq 3.9\times 10^{-2}~{\rm pc}.
\end{align} 
Notice how, for the same value of $m_a$, the characteristic scale is considerably smaller in the ALP case. 

\begin{figure}
    \centering
    \includegraphics[width=\columnwidth]{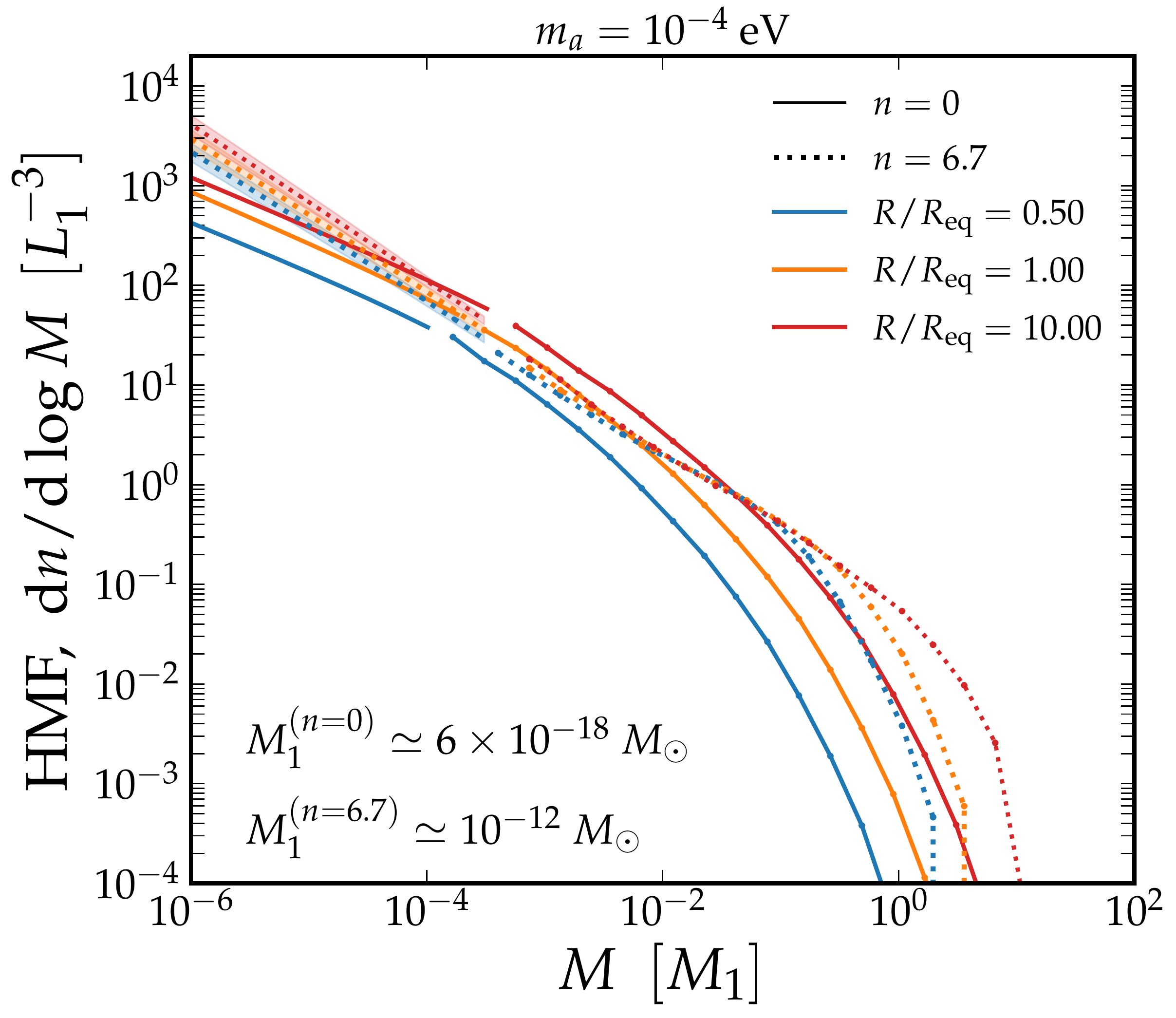}
    \caption{Comparison of the minicluster HMF at several times specified by $R/R_{\rm eq}$, where $R_{\rm eq}$ is the scale factor at matter-radiation equality. Solid lines correspond to the result for an $\n=0$ simulation and the dotted lines to an $\n=6.7$ simulation. We also label the characteristic mass scale of the miniclusters, $M_1$, for the two scenarios, noting that it is considerably small for the ALP case ($\n=0$) even while the axion mass is assumed to be the same for the two simulations at $m_a = 10^{-4}$~eV.}
    \label{fig:hmf}
\end{figure} 

Having fixed the physical scales, we can now take the density contrasts on the simulation grid at $\tau_f$ and evolve them analytically using the WKB approximation until a redshift $z=10^5$, or, equivalently, a code time of $\tau=4\times 10^{10}$ and $\tau=2\times 10^8$  for $n=0$ and $\n=6.7$ respectively. Figure~\ref{fig:hmf} shows the resulting HMF in ADM units for the $\n=0$ (solid lines) and $\n=6.7$ (dotted lines) scenarios at several different times.
Comparing the two scenarios, we see that the minicluster population is larger in the QCD axion case near the large mass cutoff as well as in the small mass limit, whereas the $\n=0$ HMF exceeds the QCD case over intermediate masses. In addition, while the QCD axion HMFs resemble those found in detailed $N$-body simulations of a similar scenario~\cite{Eggemeier:2019khm} in terms of the slopes and cutoffs, the $\n=0$ HMFs show a smoother dependence on the minicluster mass. The implication of this results, through a more accurate study of the density profiles, is remarkable for axion-like dark matter searches. Indeed, if a sizeable portion of the Milky Way’s dark matter is bound on miniclusters, the actual signal of axion dark matter
detectable by Earth-bound axion search experiments will be substantially different from conventional expectations. Furthermore, these miniclusters will likely collide with other compact objects such as white dwarfs and neutron stars,
potentially leading to observable astrophysical signature, see e.g. \cite{Edwards:2020afl,Kavanagh:2020gcy}.


Before closing this section, we would like to emphasise that strictly speaking our method of evolving the simulation grid using the WKB approximation does not capture the full dynamics of the system between $\tau_f$ and the time around matter-radiation equality.  This is especially so for the small-\n scenarios, where the group velocities of the high-$k$ modes cannot be neglected and a full treatment including the additional effects of gravitational velocities sourced by density perturbations is needed. Therefore, our conclusions on the minicluster HMF for $\n=0$, or for any small-\n scenario, should be revisited in a dedicated gravitational analysis.

\section{Conclusions}\label{sec:conclusions}

The search for the axion is ramping up: new experiments aimed at detecting these elusive particles as DM are coming online, and increasingly diverse methods to look for axionic DM indirectly are being put forward as well. Soon we will be finally probing well-motivated parameter space that has laid unexplored for decades. In order to take full advantage of this accelerating interest, it is crucial that we understand which models of axions and ALPs are consistent with our cosmological understanding of DM, and what are the present-day implications of their associated production scenarios.

In this article we have presented the results from a suite of simulations for axion-like particles in the so-called post-inflationary scenario, focusing our attention on the dynamics of the axion field around the time where its mass becomes cosmologically relevant. We have placed a particular emphasis in this study on understanding how the \emph{rate} at which this mass grows influences the ensuing dynamics, as well as the ultimate abundance and distribution of DM. Adopting the common mass-temperature parameterisation of $m^2_a \propto T^{-\n}$, we have performed simulations across the range $\n\in[0,6]$, approaching the QCD axion scenarios with $\n\sim6$--8. To our knowledge, this work includes the very first large-scale simulations of a temperature-independent ALP model ($\n=0$). As can be seen from our results, the qualitative picture of the evolution in the field---in terms of the resulting cosmic strings, domain walls, and axitons---is strongly dependent on~$\n$. In general, the faster the axion mass grows, the faster the topological defect network is destroyed, but the more small-scale structures emerge (see Fig.~\ref{fig:fullpage1} for a complete visualisation).

While the overall trends we observe in the field's evolution as we change $\n$ confirm our general expectations, the key results we have presented here are quantitative. Upon a careful inspection of the axion energy density power spectrum at the end of our simulations (see Fig.~\ref{fig:pswkb}), we find that it peaks at momenta $kL_1\in [10,50]$ corresponding to length scales much smaller than the reference value $L_1$, and the peak height increases significantly as we increase \n. This observation modifies the naive assumption that spatial fluctuations---and subsequently the seeds of miniclusters---occur on scales $\sim L_1$. 


This tendency for there to be more small-scale power in cases with faster mass growth also manifests at earlier times in the population of axitons. These are small and highly dense field configurations that appear after the destruction of the string-wall network, but before the axion mass saturates to its present-day value. We find that the density profiles of these axitons are well described by spherical Gaussian distributions of widths $\sim 1/m_a$, even for the more unstable axitons that appear in scenarios of slow axion mass growth. However, their population statistics are strongly $\n$-dependent: large values of $\n\sim 5,\,6$ lead to huge populations of axitons that are still rapidly growing at our final simulation times (see Fig.~\ref{fig:Naxitons}), whereas small values of \n result in low numbers that stabilise as the axitons emit relativistic axions.  In the case of a temperature-independent axion mass ($\n=0$), we find that axitons almost never form.

We then calculated the axion number density so as to estimate the present-day energy density in axion DM. To quantify the efficiency of the string-wall network in producing axions, we have compared the number density of axions observed in our simulations, with the density derived from an analytic nonlinear misalignment calculation (see the lower panel of Fig.~\ref{fig:dm}). We find that the smaller-$\n$ cases are the most efficient at producing axions, in agreement with the conclusions of Refs.~\cite{Klaer:2017ond,Chaumet:2021gaz}. The temperature-independent case is the most efficient of all: the presence of topological defects leads to 25\% more DM than the misalignment calculation. Somewhat counter-intuitively, for the largest-\n simulations we have performed here, which are close to predicted QCD axion values, the number density of axions is actually almost 25\% \emph{smaller} than the analytic misalignment abundance. We also realise that the estimation of the production efficiency has a substantial $\kappa$ dependence for small $\n$ scenarios, as opposed to a smaller uncertainty for QCD axion models \cite{Klaer:2017ond}.

Finally, we discussed the formation of axion-like miniclusters: gravitationally bound objects formed from the density seeds left at the end of our simulation that begin to collapse before matter-radiation equality. Using the density contrast field from our simulations, we have estimated the minicluster halo mass function following a modified Press-Schechter treatment (see Fig.~\ref{fig:hmf}). We emphasise that this is only a projected result and likely reliable only for large values of \n, as argued in Sec.~\ref{sec:gravity}. A full analysis of the gravitational evolution of these scenarios, also as a function of \n, will be the focus of our next investigations. A number of terrestrial experiments under preparation right now will be well-positioned to probe ALP DM models in the near future (see Fig.~\ref{fig:dm1}). So these followup gravitational studies will be essential if we are to understand how much the axion distribution these experiments will attempt to observe might be impacted by the presence of DM substructure predicted in the post-inflationary scenario.

\acknowledgments

We thank E.~Hardy and G.~Moore for useful comments on a draft of this article. 
GP would like to thank J.~Hamann and M.~Schmidt for discussions in the early stage of the project and M. Thompson for the technical support with the computational facilities.
CAJO is supported by the Australian Research Council under the grant number DE220100225.
GP is supported by a UNSW University International Postgraduate Award. 
JR is supported by the grant PGC2018-095328-B-I00(FEDER/Agencia estatal de investigacion) and FSE-DGA2017-2019-E12/7R (Gobierno de Aragon/FEDER) (MINECO/FEDER), the EU through the ITN Elusives
H2020-MSCA-ITN-2015/674896 and the Deutsche Forschungsgemeinschaft under grant SFB-1258 as a Mercator Fellow. 
Y$^3$W is supported in part by the Australian Government through the Australian Research Council’s Future Fellowship (project FT180100031).
This research was undertaken using the computational cluster \emph{Katana} supported by Research Technology Services at UNSW Sydney, and the \emph{Gadi} cluster from the National Computational Infrastructure (NCI), supported by the Australian Government.  We acknowledge the use of \codo{matplotlib}~\cite{Hunter:2007} and \codo{cmasher}~\cite{cmasher} for the visualisation. 

\appendix

\section{Information on the code}\label{sec:code}

\codo{jaxions} is a recently developed high-level numerical library that tracks the evolution of the axion field in the early Universe. The code is written in C++, and includes an additional Python library (\codo{pyaxions}) for efficient data analysis and visualisation. The library embodies massive hybrid parallelisation (MPI+openMP), to optimise the use of typical HPC infrastructures, especially for CPU clusters. A GPU version is currently under development. Running on the CPU, processors also make use of advanced vector extensions (AVX/AVX2/AVX512) and loops that evolve the grid in time have a cache tuner functionality to optimise the openMP chunks. 

We have tested \codo{jaxions} on the UNSW cluster \emph{Katana} and the larger NCI cluster \emph{Gadi}, comparing pure MPI runs (one CPU core per MPI process) with hybrid runs, the latter utilising the openMP multithreading as well. For large simulations an optimised parallel scaling and large-scale systems are required, since huge amounts of memory ($>200$ GB) force us to split the simulation across multiple compute nodes. A typical simulation with $3072^3$ lattice points requires $\sim 700$ GB and $\sim 4000$ CPU hours. Hence, in the running of our main simulations and tests we have used in total close to one million CPU hours.

\section{Additional visualisations}\label{sec:extra_viz}

\begin{figure*}
    \centering
    \includegraphics[width=0.9\textwidth]{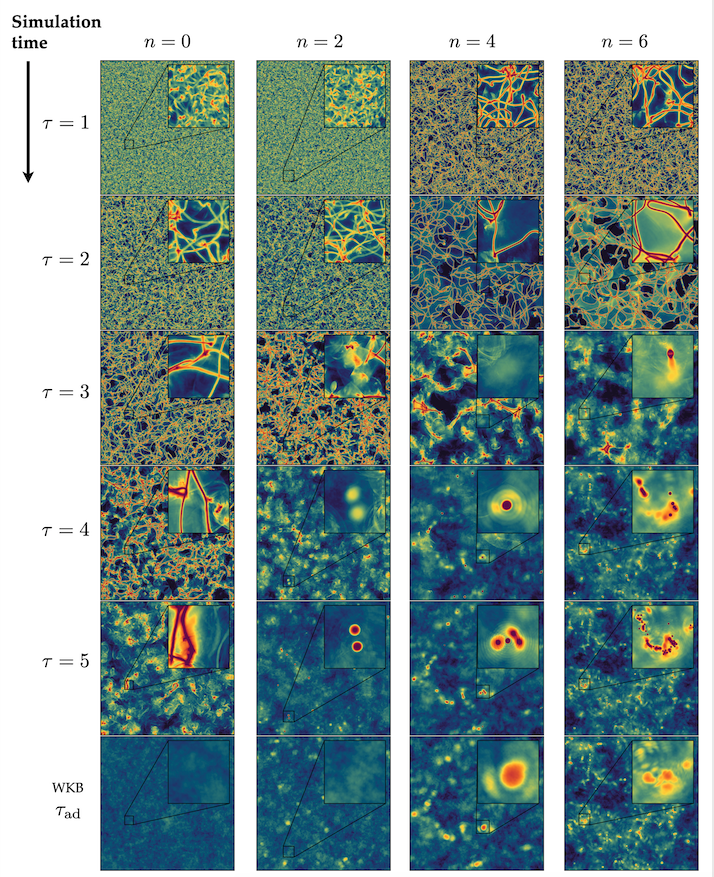}
    \caption{Evolution of the projected density contrast $\rho_a/\langle\rho_a\rangle$ as a function of simulation time $\tau$, for four values of $\n = $ 0,2,4,6. The final row also shows the projected density after the WKB approximation has been applied. The difference in timescales for the collapse of the string-wall network is apparent by comparing the panels of different $\n$ at the same $\tau$. By comparing the appearance of the field at the penultimate time $\tau = 5$, we can also appreciate the difference in the axiton population that appears as a function of $\n$, where faster mass growth leads to more plentiful axitons.}
    \label{fig:fullpage1}
\end{figure*}

Figure~\ref{fig:fullpage1} shows six snapshots of the projected axion density at different stages of evolution, including at the final time after we have applied the WKB approximation, for four values of $\n$.  In each panel we include also a zoom-in of a $250\Delta_x \times 250\Delta_x$ region of the box to give a closeup of the structure of the strings, walls, and axitons. 

\begin{figure*}
    \centering
    \includegraphics[width=0.95\textwidth]{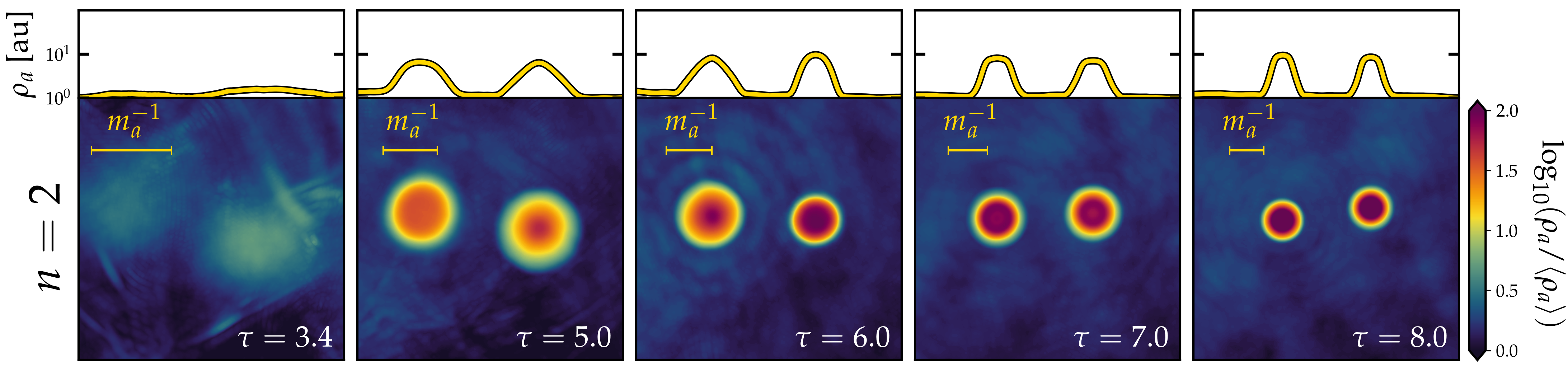}
    \includegraphics[width=0.95\textwidth]{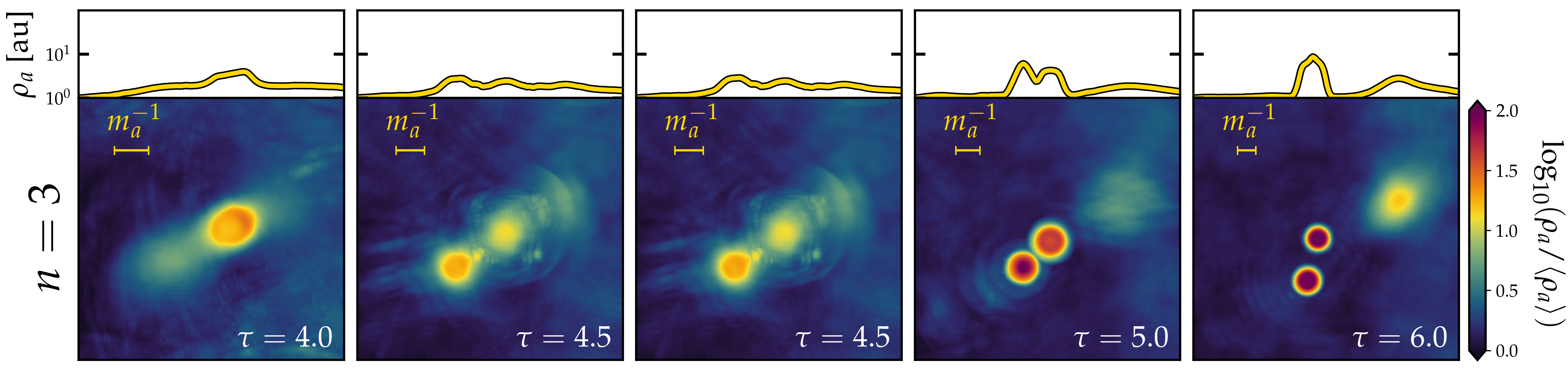}
    \includegraphics[width=0.95\textwidth]{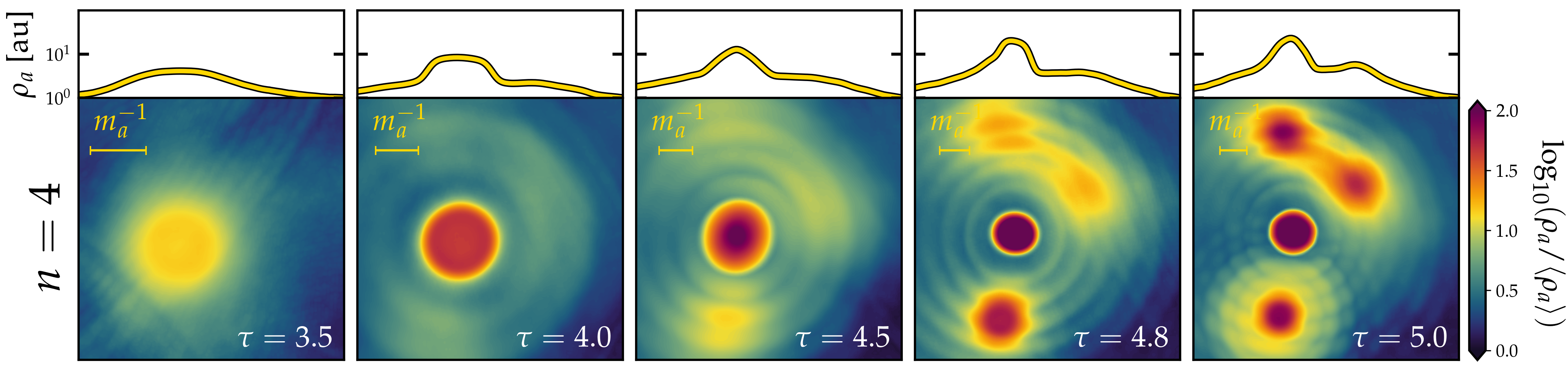}
    \includegraphics[width=0.95\textwidth]{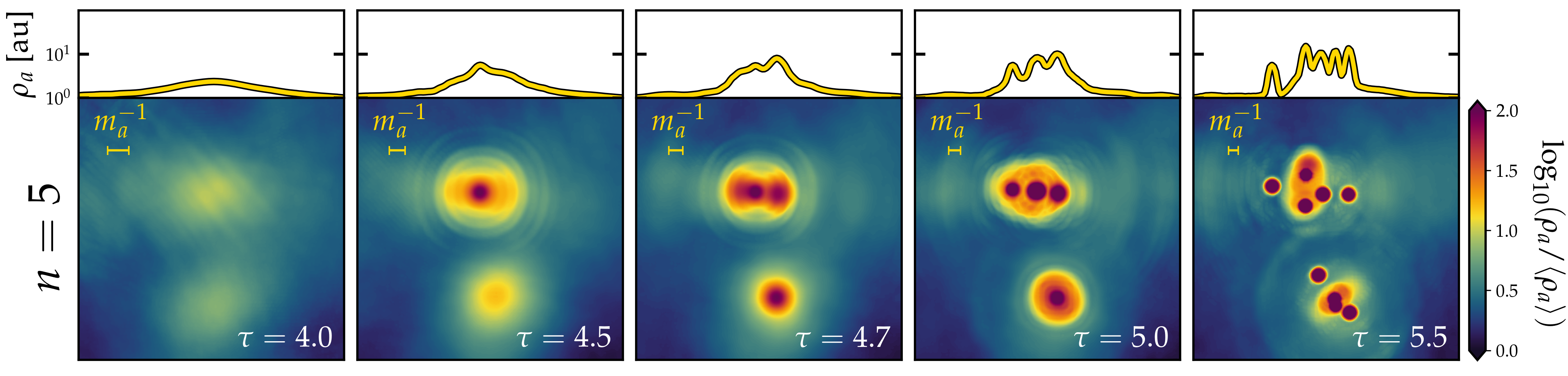}
    \caption{Same as Fig.~\ref{fig:axitonprofile_n12}, but for $\n=2,3,4,5$.}
    \label{fig:fullpage2}
\end{figure*}

Figure~\ref{fig:fullpage2}, on the other hand, shows the evolution of a $150\Delta_x \times 150\Delta_x$ region of the projected axion density as a function of time for the cases $\n= 2,3,4,5$, and has been provided here to supplement Fig.~\ref{fig:axitonprofile_n12}, which shows the same evolution but only for the  $\n = 1$ and $\n=6$ cases.


\bibliography{axions.bib}
\bibliographystyle{bibi}

\end{document}